\documentclass[11pt, a4paper, nonumbering]{deepmind}

\usepackage[authoryear, sort&compress, round]{natbib}
\usepackage{caption}
\usepackage{xspace}
\usepackage{subcaption}
\title{Towards practical reinforcement learning for tokamak magnetic control}

\newcommand{\EE}{\mathbb{E}}
\newcommand{\bad}{\emph{bad}\xspace}
\newcommand{\good}{\emph{good}\xspace}
\newcommand{\shapetask}{\emph{shape\_70166}\xspace}
\newcommand{\showcasetask}{\emph{showcase\_xpoint}\xspace}
\newcommand{\snowflaketask}{\emph{snowflake\_to\_perfect}\xspace}

\author[*,1]{Brendan D. Tracey}
\author[*,1]{Andrea Michi}
\author[*,1]{Yuri Chervonyi}
\author[*,1]{Ian Davies}
\author[1]{Cosmin Paduraru}
\author[1]{Nevena Lazic}
\author[2]{Federico Felici}
\author[1]{Timo Ewalds}
\author[1]{Craig Donner}
\author[2]{Cristian Galperti}
\author[1]{Jonas Buchli}
\author[1]{Michael Neunert}
\author[1]{Andrea Huber}
\author[1]{Jonathan Evens}
\author[1]{Paula Kurylowicz}
\author[1]{Daniel J. Mankowitz}
\author[1]{Martin Riedmiller}
\author[2,3]{The TCV Team}

\affil[*]{Equal contributions}
\affil[1]{Google DeepMind, London, UK}
\affil[2]{Ecole Polytechnique Fédérale de Lausanne (EPFL), Swiss Plasma Center (SPC), CH-1015 Lausanne, Switzerland}
\affil[3]{See the author list of \cite{reimerdes2022overview}}

\begin{abstract}
Reinforcement learning (RL) has shown promising results for real-time control systems, including the domain of plasma magnetic control. However, there are still significant drawbacks compared to traditional feedback control approaches for magnetic confinement.
In this work, we address key drawbacks of the RL method; achieving higher control accuracy for desired plasma properties, reducing the steady-state error, and decreasing the required time to learn new tasks. We build on top of \cite{degrave2022magnetic}, and present algorithmic improvements to the agent architecture and training procedure.  We present simulation results that  show up to 65\% improvement in shape accuracy, achieve substantial reduction in the long-term bias of the plasma current, and additionally reduce the training time required to learn new tasks by a factor of 3 or more. We present new experiments using the upgraded RL-based controllers on the TCV tokamak, which validate the simulation results achieved, and point the way towards routinely achieving accurate discharges using the RL approach.
\end{abstract}

\begin{document}

\maketitle

\section{Introduction}

Feedback control is vital for the operation of tokamak devices. A control system actively manages the magnetic coils to tame the instability of elongated plasmas \citep{Lazerus1990}, preventing damaging vertical disruption events \citep{lehnen2015disruptions}. Furthermore, precise control of the plasma current, location, and shape enables management of heat exhaust and plasma energy \citep{silburn2017mitigation, leonard2005divertor}. In research contexts, scientists study the effects of changes in the plasma configuration on these quantities of interest \citep{moret1997,hofmann2001stability,anand2017,anand2020plasma, pesamosca2021model}, requiring control systems for novel configurations and rapid variation around the nominal scenario. A flexible toolset that enables rapid iteration and delivers precise magnetic control is thus a boon to tokamak research and development.

Traditionally, accurate control of the plasma is achieved through successive loop closures of the plasma current, shape, and location \citep{de2019plasma}. In this paradigm, the control designer pre-computes a set of feedforward coil currents \citep{hofmann1988fbt,blum2019automating}, and then builds feedback loops for each of the controlled quantities. These quantities (e.g. plasma shape and location) cannot be directly measured, and must be indirectly estimated in real-time from magnetic measurements. In particular, the shape of the plasma must be estimated in real-time using equilibrium reconstruction codes \citep{ferron1998real, moret2015tokamak, blum2008}. Such systems have successfully stabilized a wide range of discharges, but design can be challenging and time-consuming (in researcher and experimental time), especially for novel plasma scenarios.

Reinforcement learning (RL) has recently emerged as an alternative paradigm for building real-time control systems. While historically high-profile successes of reinforcement learning were confined to simulated systems such as StarCraft \citep{vinyals2019grandmaster} and DOTA \citep{berner2019dota} or highly structured systems such as Go \citep{silver2017mastering}, more recent successes include systems with high amounts of hidden knowledge like Stratego \citep{perolat2022mastering} and Diplomacy \citep{meta2022human}, and large unstructured spaces such as language \citep{christiano2017deep, ouyang2022training}. RL is also being deployed for increasingly sophisticated real-world applications such as data center cooling \citep{luo2022controlling} and memory management \citep{wang2023optimizing}.

Alongside these more general successes, reinforcement learning is becoming increasingly used for plasma control. While early works focused on supervised learning, for example to perform real-time estimation of coarse plasma shapes \citep{Lister1991,bishop1995real}, recent efforts have expanded to using RL, for example to construct feedforward trajectories of plasma parameters such as $\beta$  \citep{seo2021feedforward, seo2022development}, control of $\beta_N$ \citep{char2020offline, char2021model, char2022differential}, and direct control of the vertical instability \citep{dubbioso2023}. For a recent overview of ML applications to fusion research, including the use of RL, please see \citep{pavone2023machine}.

Recent work by \cite{degrave2022magnetic} demonstrated the ability for an RL-designed system to perform the main functions of tokamak magnetic control. In particular, this work presented a system where an RL ``agent'' learns to control the Tokamak à Configuration Variable (TCV) \citep{hofmann1994creation, reimerdes2022overview} by interacting with the FGE tokamak simulator \citep{carpanese2021development}. The control policy learned by the agent was then integrated into the TCV control system, whereby the policy observed TCV's magnetic measurements, and output control commands for all 19 magnetic control coils. \cite{degrave2022magnetic} demonstrated the capability for RL agents to control a wide variety of scenarios, including plasmas that are highly-elongated, snowflakes \citep{anand2019real}, and even demonstrated a novel stabilization of a ``droplet'' configuration with two separate plasmas in the vacuum chamber simultaneously. This work presents a strong case for RL-designed control systems, where the control designer expresses a final goal (quantified using a reward function) that is maximized by the agent. This shifts the focus away from the exact specifics on ``how'' to achieve such goals, and toward ``what'' goals should be achieved. 

RL approaches, however, have a number of drawbacks which have limited their uptake as a practical solution for the control of tokamak plasmas. Here we address and begin to alleviate three of these challenges: the difficulty in specifying a scalar reward function that is both learnable and provokes accurate controller performance; the steady-state bias in tracking errors; and long training times.
First, in \nameref{Reward Shaping}, we propose a method for reward shaping as an intuitive and simple solution to improve control precision. We then address the issue of steady-state error, in \nameref{Integrator Feedback}, by providing explicit signals for error and integrated-error to the agent. This reduces the accuracy gap between classical and reinforcement-learned controllers. Finally, in \nameref{Episode Chunking} and \nameref{Transfer learning}, we address the issue of training time required to generate control policies. RL algorithms are known to have high computational cost and low sample efficiency \cite{cabi2019scaling}, a problem exacerbated for tokamaks where even low-fidelity plasma simulators are significantly more computationally expensive than simulators used in traditional RL applications. We address this by using a multi-start approach for complex discharges and show substantial reductions in the training time of new policies. Furthermore, we show that warm-starting training with existing control strategies can be a very effective tool when the new scenario of interest is close to a previous scenario. In combination, these techniques lead to a significant reduction in training time and improvement in accuracy, making substantial strides towards enabling RL to be a routinely usable technology for plasma control. 

\section{Background}

\subsection{Reinforcement Learning}

Reinforcement learning \citep{sutton2018reinforcement} is a subset of machine learning that represents control problems as the interaction between an \emph{agent} and an \emph{environment}. The agent can be seen as the control mechanism, and the environment can be seen as the system to be controlled.
The agent receives a set of signals, called \emph{observations}, and sends a set of control signals, known as \emph{actions}.
In standard RL settings, the agent-environment loop operates at discrete intervals. The state of the environment at time $t$ is denoted by $s_t$, and the measurements observed are a function of this state $o_t = o(s_t)$ . The agent chooses actions according to a control strategy, also known as a \emph{policy}, which is a potentially non-deterministic function $a_t = \pi(o_t)$. The environment is influenced by the action taken, and evolves according to a (potentially non-deterministic) dynamics function $s_{t+1} = s(s_t, a_t)$. In traditional control techniques, the goal is generally to keep the error small between a desired reference value and the actual measured value (or measurement-estimated state). Reinforcement learning, in contrast, works with the more general concept of \emph{reward}. While optimal control also uses the concept of reward (or cost respectively), practical algorithms typically force restrictions on the formulation of the system and/or reward. The reward function $r_t = r(s_t, a_t, s_{t+1})$ is a real-valued function whose output indicates the quality of executing action $a_t$ in state $s_t$, where higher is better. Note that in general the reward is a generic function of the environment state and action, though often rewards will just depend on $s_{t+1}$, the state reached as a result of the action. While it can relate to an error signal (with increasing reward for decreasing error), it can also be a function combining multiple different error signals, or include terms that penalize certain actions or states (e.g. control coil currents being too large).

An \emph{episode} of interaction between the agent and the environment begins with the environment in an initial state $s_0$. The agent observes $o_0 = o(s_0)$, and takes an action $a_0 = \pi(o_0)$. The environment then advances state to $s_1 = s(s_0, a_0)$, and the agent receives a reward of $r_0$. The agent then takes action $a_1$, the environment proceeds to $s_2$, and the agent receives $r_1$. This continues iteratively until a termination is triggered (e.g. a limit of $T$ steps, or an off-nominal condition). In simulation, this discrete time approximation is natural, though of course is only an approximation to the continuous time environment evolution in physical experiments. The goal in RL (and of RL algorithms) is to find the optimal policy to maximize the discounted accumulated reward over an episode when starting from an initial state $s_0$:
\begin{equation}
\pi^* = \arg \max_\pi \EE^{\pi} \left[ \sum_{t=0}^T \gamma^t r(s_t, a_t, s_{t+1})\  \right],
\end{equation}
where $0 < \gamma \leq 1$, where $\gamma$ is a discount factor controlling the myopia of the agent.

\subsection{RL for Plasma Magnetic Control}
\label{RL for Plasma Magnetic Control}

We follow the work of \cite{degrave2022magnetic} in translating the challenge of magnetic control on TCV into a reinforcement learning problem. During a discharge, the environment state, $s_t$, which for the present purpose we can consider as the complete state of the tokamak, includes the plasma, the electric currents in the control coils and passive structures, and derived quantities such as the resulting magnetic field (as will be discussed, this state is simplified in simulation). The agent acts at a $10$kHz control rate, sending a voltage command to each of the 19 magnetic control circuit voltage power supplies at every time step (2 ohmic circuits, 8 each of high-field-side and low-field-side poloidal field coils, and one in-vessel vertical control circuit consisting of up-down antiseries-connected windings).

The agent observation, $o_t$, consists of two components: the real-time sensor measurements and the control targets, known as the \emph{references}. There are $92$ real-valued sensor measurements; $34$ wire-loop magnetic flux measurements, $38$ magnetic field probes, $19$ control coil current measurements, and one (redundant) measurement of the difference in currents between the ohmic coils. The references represent the desired plasma configuration, including the location of the plasma, the limit point, and any desired X-points. 
The plasma boundary is defined by the last closed-flux surface (LCFS), which is the outermost closed surface of the iso-flux contours. The desired LCFS is represented in the references by the $r$ and $z$ coordinate locations for $32$ control points. The references may additionally include explicit targets for properties of the LCFS, such as reaching a desired elongation or triangularity. X-points \citep{grad1958hydromagnetic} are saddle points of the magnetic flux corresponding to points with $B_r = B_z = 0$, where $B$ is the magnetic field, and again, the desired X-point locations are represented with their $r$ and $z$ coordinates. The limiting point is also represented with an $r$ and $z$ coordinate, either a location along the limiter for limited plasmas, or the location of the active X-point for diverted plasmas. Finally, we additionally provide a desired value for the plasma current, $I_p$.

\begin{figure}[ht!]
\centering
\includegraphics[width=0.68\linewidth]{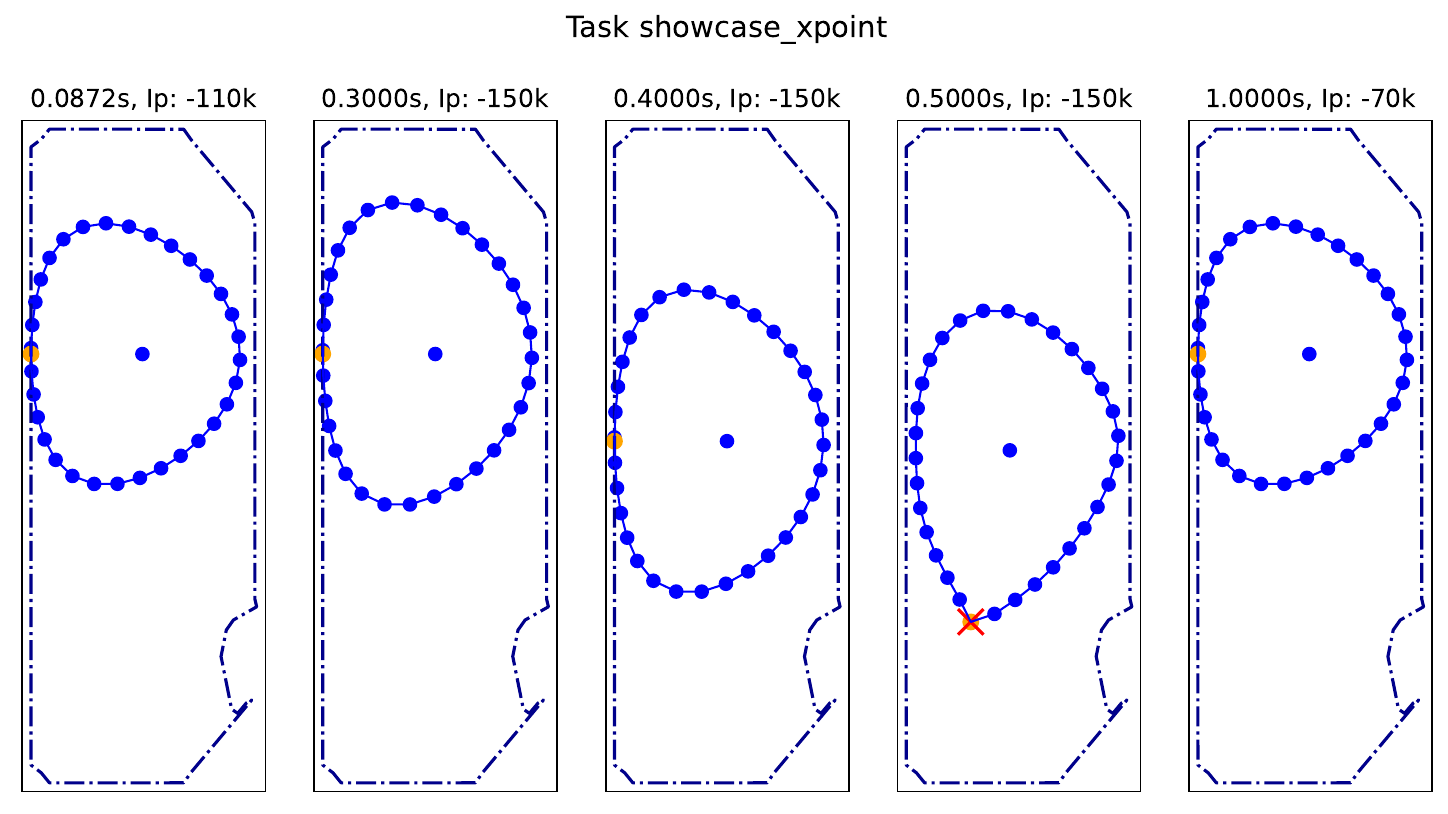}
\caption{A simplified set of references (desired plasma configurations) for the \showcasetask task. Time (measured from the start of the discharge) and the plasma current ($I_p$) are displayed above each plasma shape. X-points marked with red "\textcolor{red}{X}". The limit point is displayed in \textcolor{orange}{orange}. For the full reference set, see Fig. \ref{fig:showcase_xpoint_references_all}.}
\label{fig:showcase_xpoint_references}
\end{figure}

For most tasks, the references are not static, but instead a function of time. We pre-determine a desired evolution of the plasma during the discharge, an example of which is shown in Fig. \ref{fig:showcase_xpoint_references}. During execution of the experiment, the measurements and references at time $t$ are combined together to create the observation, $o_t$, from which the voltage commands are determined. Note that the RL agent receives the raw magnetic measurements, not the direct state or a reconstruction of it. No explicit error measurements are provided to the agent, and no observers are needed to convert measurements into error values.

In this work, we use the same basic experimental design as in \citet{degrave2022magnetic}. We learn a control policy, $\pi$, for a specific experiment through interaction with a simulated environment, and then deploy the resulting policy for a discharge on TCV. Specifically, the dynamics are modeled with the free-boundary simulator, FGE, with additional stochasticity added to model the noise in sensor values and power supplies, and to vary the parameters of the plasma. The sensor noise is applied per environment step, while the plasma parameter variation (the plasma resistivity, $R_p$, the normalized plasma pressure, $\beta_p$, and the safety factor at the plasma axis, $q_A$) is simplified so the values are constant within an episode, but are randomly sampled between episodes.

Like \cite{degrave2022magnetic}, we use the Maximum-a-Posteriori Optimization (MPO) algorithm \citep{abdolmaleki2018maximum} to develop control policies. MPO relies on two neural networks: an ``actor'' network that outputs the current policy $\pi$, and a ``critic'' network that approximates the expected accumulated reward of that policy. The agent interacts with $1000$ copies of the FGE environment, collecting the observations seen, the actions taken, and the reward received.
The reward received at each step is computed based on how close the plasma state is to the target values contained in the references, augmented by other factors such as avoiding undesirable plasma states. A straightforward translation from the optimal control paradigm to reinforcement learning would be to have a reward component for each error term to be minimized, where each component $i$ is mapped to a scalar value $x_i$. These values are then combined into a single scalar reward value. Based on recorded sequences of observations, actions, and rewards, the agent alternates policy updates and critic updates using gradient descent on a regularized loss function (see \citet{abdolmaleki2018maximum} and the \nameref{Reward Shaping} section for details). The updated actor network parameters are used in future interactions with the environment. For plasma discharges, the actor network is restricted to a small architecture that can execute at $10$kHz, but the critic network is only used during training, and so can be sophisticated enough to learn the environment dynamics (which the actor network then uses to learn what is needed for control). The major implementation differences with the setup in \cite{degrave2022magnetic} and the work presented here are that:
\begin{itemize}
  \item All experiments in our setup use JAX \citep{jax2018github} and Haiku \citep{haiku2020github} for neural network construction and parameter training.
  \item The error tolerance on the residual of FGE's Newton-Krylov solver was reduced from 1e-8 to 1e-4. This lead to a training speedup of 2-3x without significantly affecting the simulated control outcome.
\end{itemize}

As noted above, we can define a \emph{task}, i.e. desired evolution of the plasma state during a discharge, through a time-varying set of references. In this work, we consider the following tasks:
\begin{itemize}
    \item \shapetask: a simple stabilization task where the plasma state is maintained as a low-elongation limited plasma. The reference values are constant throughout the episode.
    \item \showcasetask: a time-varying discharge where the agent needs to elongate the plasma and ramp up the plasma current, then shift its vertical position, then divert the plasma (create an active X-point), and finally ramp down the plasma current and reduce the elongation. Fig. \ref{fig:showcase_xpoint_references} illustrates a subset of states for this task. For all references see Fig. \ref{fig:showcase_xpoint_references_all}. This corresponds to the task in Fig. 2 of \cite{degrave2022magnetic}. 
    \item \snowflaketask: the snowflake configuration \citep{ryutov2007geometrical} consists of a hexagonal saddle-point structure (second order null) by placing two X-points closely together. In this task, we initially establish these two X-points at a distance, and then bring them together. The reference is shown in Fig. \ref{fig:snowflake_to_perfect_references}. This corresponds to the task in Fig. 3d of \cite{degrave2022magnetic}. 
\end{itemize}

\section{Towards Practical RL Controllers}

In this section, we detail the contributions of this paper with regard to the agent training process. First, we discuss improving the control accuracy through reward shaping. This is followed by our work toward reducing the steady state error through integral observations. We then discuss episode chunking which is used to improve wall clock training time. Finally we explore transfer learning as a means to improve training time.

\subsection{Reward Shaping}
\label{Reward Shaping}

Where traditional control algorithms take actions to minimize the error of an actively measured (or estimated) quantity, RL algorithms instead seek to maximize a generically-defined reward signal.
This reward maximization objective drives the evolution of the agent's behavior during training. However, the reward value is not computed during deployment.

In classical control algorithms, the performance of the controller can be adjusted by explicitly tuning control gains (e.g. to modify responsiveness or disturbance rejection) and adjusting trade-off weights for multi-input multi-output (MIMO) systems.
By contrast in RL, the reward function is of central importance to the learned controller's behavior. Careful design of the reward function is therefore necessary to adjust controller behavior.
In this section, we explore how modifying the design of the reward can elicit desired behaviors in the resulting trained agent.
We will see that, by adjusting the design of the reward function, we can quickly adapt agent behavior and trade off elements of our objective.
Moreover, we demonstrate that shaping the reward function is essential for creating accurate RL control policies.
We further show that it is possible to adapt an agent to a new objective by continuing training with an updated reward function.

\subsubsection{Reward Design Overview}
We modify the reward function designed for magnetic control by \cite{degrave2022magnetic} (detailed in Tables 3, 4 and 5 of their manuscript). The reward function is a combination of separate components, and
in this section, we consider how the design of these components can be altered to influence the behavior of the trained agent, in particular where there are trade-offs between performance across reward components.

Reward components correspond to different desiderata of an ideal agent (accurate shape, accurate plasma current, etc.). 
Each reward component is calculated by taking the difference between the desired value and the value reported by the simulated environment. A non-linear scaling and transformation is applied to this difference, giving an effective reward for that component. The overall (scalar) reward is computed using a non-linear combination of the individual component rewards. It is in designing the reward components where we have the most fine-grained control over incentives for the agent. 

In this work, we combine the reward component values using a weighted \texttt{SmoothMax} function. In some cases, an individual reward component is built from several related error quantities, such as the shape error at multiple control points. We also utilise the \texttt{SmoothMax} function to combine these errors into a single scalar component reward. The definition of the \texttt{SmoothMax} function is provided below.

\begin{equation}
\texttt{SmoothMax}(x_1, x_2, ..., x_n, w_1, w_2 ..., w_n, \alpha) = \frac{\sum_{i=1}^n w_i x_i e^{\alpha x_i}}{\sum_{i=1}^n w_i e^{\alpha x_i}}.
\end{equation}
The choice of the $w_i$ directly provides the (relative) importance of each component which itself is carefully designed. The value $\alpha$ affects the trade-offs between ``easy'' and ``hard'' (to satisfy) components. A value of $\alpha$ much less than zero (say, $\alpha < -5$) means that the reward received by the agent is nearly equal to the component it is performing least well on, while a value of $\alpha$ 
close to zero means that all components are equally emphasized. A value of $\alpha$ greater than zero is unsuitable, since it would accentuate components that are controlled well at the exclusion of components where it is doing poorly.

Many individual components that feed in to the \texttt{SmoothMax} are constructed similarly to those of classical controllers (e.g. keep the plasma current close to a desired quantity). However, reward components are not constrained to be (easily) measurable from sensor measurements, providing additional flexibility in their construction. Reward components can also be multi-modal, for example to encourage behavior away from regions of state-space that are undesirable or less well modeled by the simulator.

\begin{figure}[ht]
\centering
    \begin{subfigure}{0.49\linewidth}
        \centering
        \includegraphics[height=0.2\textheight]{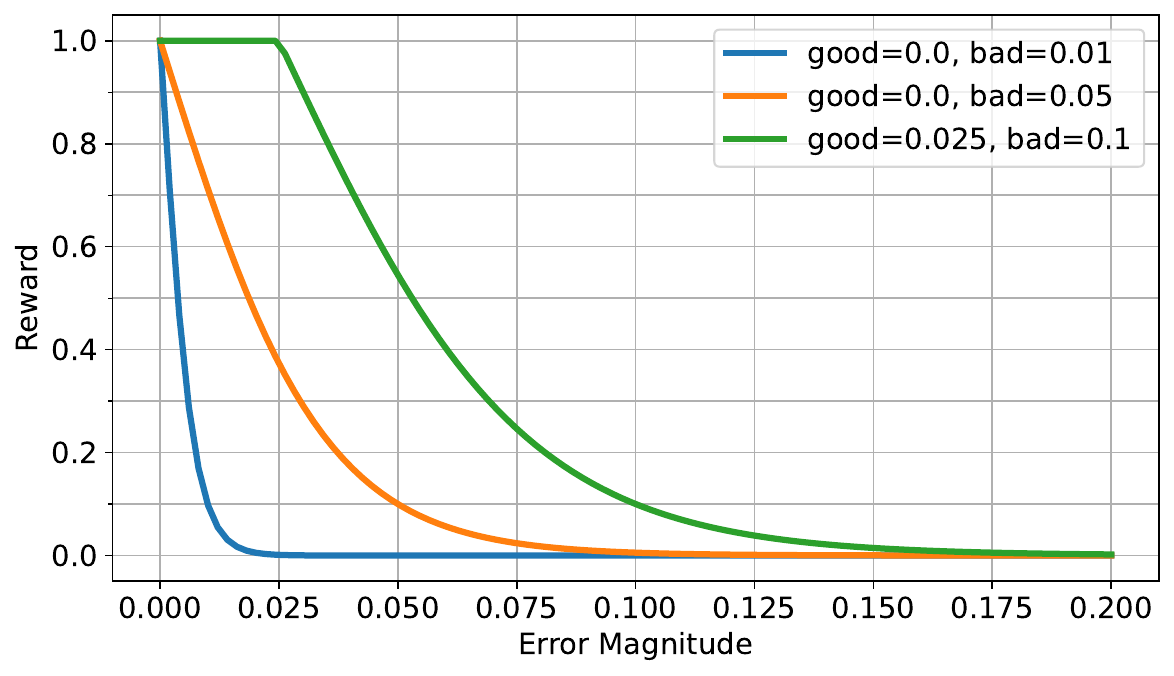}
        \caption{Plot depicting the impact of the \good and \bad reference values on reward component values.
        %Both the rewarding region and the gradient of the reward value with respect to error magnitude are affected.
        Note that the \bad reference points correspond to a reward of 0.1.}
        \label{fig:reward_parameter_influence}
    \end{subfigure}
    \hfill
    \begin{subfigure}{0.45\linewidth}
        \centering
        \includegraphics[height=0.2\textheight]{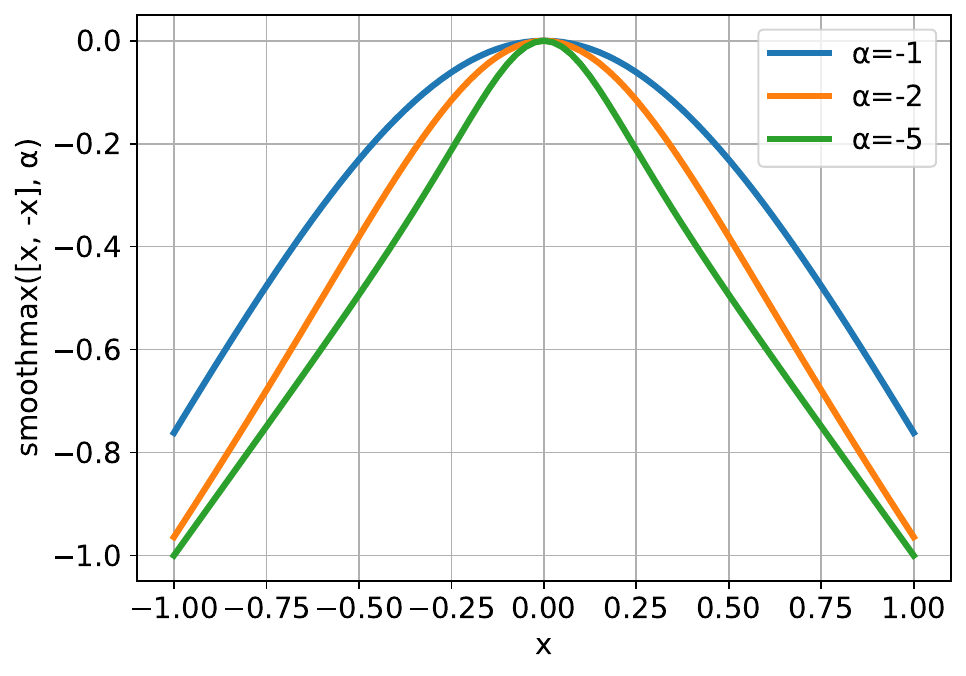}
        \caption{$\texttt{SmoothMax}(x_1=x, x_2=x, w_1=1, w_2 = 1, \alpha)$ \vspace{2.2em}}
        % 
        % }
        \label{fig:smoothmax_demo}
    \end{subfigure}
\caption{Visualisation of the influence of reward hyperparameters.}
\label{fig:reward_design}
\end{figure}

In this work, we use a \texttt{SoftPlus} transformation to arrive at scalar reward components:\footnote{Our \texttt{SoftPlus} implementation is based on the lower half of the logistic function instead of the standard \emph{SoftPlus} since we want it to be bounded from 0 to 1, with the \good value being exactly 1.}
\begin{equation}\label{eq:softplus}
\begin{aligned}
    \texttt{SoftPlus}(x) &= \left[2\cdot  \sigma\left(f_\text{scale}(x, \texttt{good}, \texttt{bad}, 0, \zeta\right)\right]_0^1,\\
    f_\text{scale}(x, u_1, l_1, u_2, l_2) &= u_2 - \left(\frac{x - l_1}{u_1 - l_1}\right)(u_2 - l_2),\\
    \sigma(x) &= \frac{1}{1 + \exp(-x)},\\
    [x]_a^b &= \min\left(\max(x, a), b\right).
\end{aligned}
\end{equation}
The \good and \bad parameters act to scale the reward signal into a region of interest. If the true value is worse than \bad, the reward rapidly decays toward a value of 0,
while if it is at or better than \good, the reward saturates to a value of $1$. The parameter $\zeta$ affects the ``sharpness'' of the reward scaling between the \good and \bad reference points\footnote{In our experiments we set $\zeta = -\log(19) = -2.9444$ such that the \bad reference point corresponds to a reward of 0.1.}.
The set of scalar reward components are then combined with a (weighted) \texttt{SmoothMax} operator to attain a final scalar reward. These functions as well as the influence of hyperparameter choices is depicted in Figure \ref{fig:reward_design}.

Components with tight \good and \bad parameters must be controlled well to receive a high component reward, while components with looser constraints are more easily satisfied.

In theory, many parameter choices should be approximately equivalent as they are monotonic adjustments to the reward and should not strongly affect the optimal policy. In practice, however, we are reliant on gradient descent, and do not have a perfect global optimizer. We need to deal with exploring a global space in the face of stochastic returns.  Tight values of \good and \bad make it difficult to find a region to get any appreciable reward (or an appreciable gradient on how to improve). On the other hand, a loose value of \bad makes it easier to find a reward signal, but harder to discover precise control as there is a smaller change in reward upon improvement. Intuitively, ``tight'' reward parameters may therefore be more appropriate where the initial conditions are close to the goal state and thus the reward does not need to shape goal discovery as well as fine-grained accuracy.

Relatedly, a value of $\alpha$ near zero will provide a reward gradient on all components, but provides a more diluted signal on all components. In contrast, a substantially negative value of $\alpha$ will provide a strong incentive to improve the worst component, but takes away incentive to improve other components and thus harms exploration. These trade-offs are especially prevalent in the plasma domain where reward components are often complementary (or at least orthogonal) rather than truly being in contrast, e.g. accurate X-point control assists in accurate LCFS control, and does not detract from accurate plasma current control.

\subsubsection{Reward Shaping in a Simple Setting}
For our initial experiments, we consider three training approaches focusing on minimizing shape error by modifying the hyperparemeters of the reward component for shape error in the \shapetask task:
\begin{enumerate}
    \item \textbf{Baseline}: the default reward parameters from \cite{degrave2022magnetic} - \good = 0.005, \bad = 0.05.\footnote{The base reward is formed from the Plasma Current, E/F Current, OH Coil and Plasma Shape components from the ``Elongated Plasma'' reward used by \cite{degrave2022magnetic} (see Extended Data Table 3 of their manuscript) with the shape component parameters provided here.}
    \item \textbf{Narrow Reward}: updating the parameters to  \good $= 0$ and \bad$= 0.025$. These reference values produce a more exacting reward function. This concentrates the reward signal at lower error values and offers a guiding signal even for small error values, providing incentive for increased accuracy in controlling the shape.
    \item \textbf{Reward Schedule}: scheduling the values of \good and \bad to gradually become more peaked as training progresses, with \good $= 0$ and  \bad decreasing from $0.1$ to $0.025$ over 6 million policy update steps. This schedule provides a wider reward basin at the beginning of training to help exploration, gradually tightening the reward to encourage accuracy as training progresses. Historical data is not relabelled as the reward function evolves; however, stale data does eventually drop out of the learning agent's replay buffer.
\end{enumerate}

\begin{table}[]
\centering
\begin{tabular}{|l|l|l|}
\hline
\textbf{Experiment}  & $\boldsymbol{I_p}$ \textbf{Error (\%)} & \textbf{LCFS Mean RMSE (cm)}  \\
\hline
Baseline & 0.353 $\pm$ 0.221 & 0.567 $\pm$ 0.221 \\
Narrow Reward & \textbf{0.238} $\pm$ 0.076 & \textbf{0.201} $\pm$ 0.057 \\
Reward Schedule & 0.450 $\pm$ 0.321 & 0.490 $\pm$ 0.196 \\
\hline
\end{tabular}
\caption{Results from reward adjustment in the simple stabilization task, \shapetask. The values are averages over 5 training runs with the the 95\% confidence interval indicated. The $I_p$ target is -120kA.
}
\label{table:reward_shaping_results}
\end{table}

The results from this first set of experiments, shown in Table \ref{table:reward_shaping_results}, demonstrate that the reward chosen for training can have significant influence over the performance of the final trained agent. Focusing on the shape error, we note that the greatest impact came from the Narrow Reward with its highly demanding static reward function. In this simple task, the more precise reward function provides a strong incentive for the controller to improve accuracy. While such a sharp reward signal can harm policy discovery, as discussed above, the task is to maintain the handover position, and so exploration is not a strong challenge in this task. With little need for exploration to find the highly rewarding states, the agent can focus on satisfying the demanding reward signal. Furthermore, the simplicity of the task means that there is little-to-no trade off in accurate control between the reward components.

The agent trained with the Reward Schedule, progressively reducing the value of \bad, shows no significant improvement over the baseline agent in terms of the targeted reward component (LCFS RMSE). We hypothesize that the poor performance of the agent trained with a reward schedule, as compared to one trained with a fixed reward, is due to the non-stationarity that a changing reward function induces during the more exploratory phase of learning.
The agent is able to learn a `good enough' policy with the initial loose reward function. The value function which estimates the expected sum of future rewards, which is in turn used to update the policy, struggles to stay in line with the latest reward function as it evolves. In essence, the value function is receiving different learning signals for the same states at different points in training without any explicit engineered ability to adapt to this. This therefore makes the learning problem much more challenging. We note this behavior despite the schedule becoming stationary at approximately 60\% through training and therefore the final 40\% of training is with a fixed reward function. An additional factor to consider is that the Gaussian noise added to the actions during training generally decreases throughout training as is natural under the MPO learning algorithm \citep{abdolmaleki2018maximum}. This leads to less exploration as training continues and therefore a potentially reduced capacity to adapt to an evolving reward as training continues.
The results from this simple setting of maintaining and refining the plasma's handover shape, suggest that a more exacting reward can improve agent performance. 

% [Shaping for different criteria]
\subsubsection{Reward Shaping for Complex Tasks}
We now turn to the \snowflaketask task where training is more costly and the reward-tuning more complex, due to time-varying objectives and a greater number of metrics of interest. In particular, we seek to improve X-point location accuracy through reward shaping.

We consider the following reward shaping approaches for X-point location accuracy:
\begin{enumerate}
    \item \textbf{Baseline}: train with default parameters taken from \cite{degrave2022magnetic} --- \good = 0.005, \bad = 0.05.
    \item \textbf{X-Point Fine Tuned}: first train with default parameters and then perform a second phase of training with a more exacting reward which emphasizes X-point accuracy --- \good = 0, \bad = 0.025.
    \item \textbf{Narrow X-Point Reward}: train with a more exacting reward function from the inception of training --- \good = 0, \bad = 0.025.
    \item \textbf{Additional Training}: perform the additional phase of training without updating the reward. This allows us to distinguish between effects from more training and changing the reward function.
\end{enumerate}

\begin{table}[]
\centering
\begin{tabular}{|l|l|l|l|}
\hline
\textbf{Experiment} & $\boldsymbol{I_p}$ \textbf{Error (\%)} & \textbf{LCFS Mean RMSE (cm)} & \textbf{X-point Location Error (cm)} \\ 
\hline
Baseline & 0.848 $\pm$ 1.710 & 1.122 $\pm$ 1.460 & 0.669 $\pm$ 0.491 \\ % & -\\
X-Point Fine Tuned & 0.717 $\pm$ 0.624 & 0.845 $\pm$ 0.097 & \textbf{0.289} $\pm$ 0.027 \\ % & -\\
Narrow X-Point Reward & 6.143 $\pm$ 4.602 & 4.536 $\pm$ 3.268 & 1.199 $\pm$ 1.102 \\ % & -\\
Additional Training & \textbf{0.502} $\pm$ 0.423 & \textbf{0.723} $\pm$ 0.159 & 0.541 $\pm$ 0.112 \\ % & -\\
\hline
\end{tabular}
\caption{Results from reward adjustment in \snowflaketask task. The baseline agent was trained with the reward used by \cite{degrave2022magnetic}. This agent was then used as a base for the additional training agent (which was allowed to continue training under the same conditions) and the fine-tuned agent which continued training with an updated reward emphasizing X-Point location accuracy. The Narrow X-Point Reward agent was trained with the shaped reward from scratch. The values are averages over 5 training runs with the the 95\% confidence interval indicated.}
\label{table:reward_xp_finetuning_results}
\end{table}
We compare the performance of four differing training configurations listed above.
The results are summarized in Table \ref{table:reward_xp_finetuning_results}.

This set of experiments demonstrate the value of a flexible reward specification. We see in the X-Point Fine Tuned result that this two-stage training procedure emphasizing X-point accuracy leads to a 57\% reduction in X-point location error. Note this is not merely due to the extra data, we see a significant improvement in X-point accuracy when compared with the Additional Training experiment. This demonstrates the importance of reward shaping rather than extra training cycles in this result. Note, however, that if we compare the X-Point Fine Tuned agent with the Additional Training agent, we see that accuracy on other quantities, such as $I_p$ and shape error, degrades slightly -- the fine-tuned agent performs less well in these respects compared to the agent that continued training with the original reward. While these differences are not statistically significant, they suggest a possible trade off that arises from reward shaping. In such cases, a control designer can assess the trade-offs of shaping different components and search for the most desirable policy.
Reward shaping acts as an intuitive tool for performing this policy search in the space of reward functions.

The agent trained with the more exacting reward throughout training (Narrow X-Point Reward) fails to learn an effective policy. Investigation of the agent's performance shows that the agent typically struggles to successfully establish the snowflake shape, failing to bring in the second X-point at the desired location.
The increased sharpness in the X-point reward causes the region where the agent receives appreciable positive signal to decrease, so when a second X-point is brought in incorrectly the agent receives a low reward, and also low signal on how to improve. Note that the standard deviation is quite high --- it is possible for the exploration to succeed and the learning to do well, but the narrow reward decreases the probability of finding the right local minimum. 
This is compounded by the low value of $\alpha$
which emphasizes the least performing element, and thus further disguises policy improvements. For example, improving LCFS accuracy might help the agent discover how to divert the plasma and thus control its limiting by the X-point location. However, a harsh (i.e. large magnitude) $\alpha$ and an exacting X-point location reward obfuscates this progress since reward calculation is dominated by X-point accuracy for which the agent receives little to no signal on how to improve.

The failure of the agents trained with the Narrow X-Point Reward from scratch is contrasted to the success of the agents trained with a more exacting shape error component in the simpler \shapetask task. This demonstrates the importance of matching the design of the reward not only to the desired agent policy characteristics but also the task at hand. In the case of \shapetask, there is no exploration needed to find the highly rewarding region of exacting reward components since the initial state is the one to be stabilized and persisted. In contrast, the \snowflaketask task requires manipulating the plasma through a series of target shapes and locations. Therefore, learning from scratch with an exacting reward is difficult. This trade-off motivates our explorations into reward scheduling and agent fine-tuning. It must be stated, however, that reward function design is an art, and there may be a better structure that maintains flexibility while mitigating this trade-off. 

Our results from the \snowflaketask task demonstrate the benefits from a two-stage training regime, training initially on a more forgiving reward and switching to a more exacting reward. We have demonstrated this using RL for both stages, though one could, for example, use RL to find the right ``local-minimum'' for the control strategy, and then in the second stage use an optimal-control based algorithm to fine-tune on the more exacting reward.
%   e
This two-stage training regime also suggests the potential of a general base controller to be rapidly adapted to a precisely defined set of targets.
This in turn is an intermediate step towards a general \textit{and} precise controller for a multitude of plasma control tasks.

\subsection{Integrator Feedback}
\label{Integrator Feedback}

\begin{figure}[ht!]
        \includegraphics[width=0.5\linewidth, trim={0 0 0 0},clip]{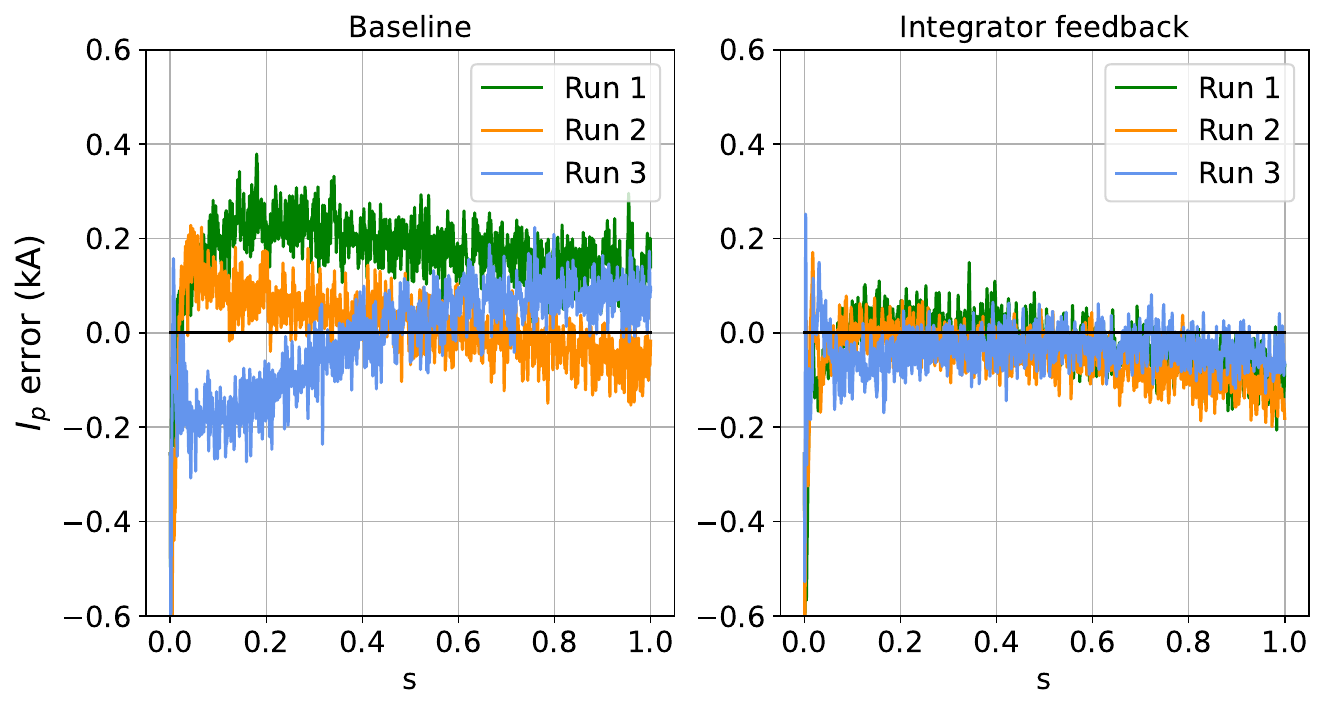}
        \includegraphics[width=0.5\linewidth, trim={0 0 0 0},clip]{figures/ss_shape_error.pdf}
\caption{Simulated errors in the plasma current and shape on the \shapetask task over a 1 second (s) control window. We compare policies with and without the average-error feedback for three random seeds. The figures demonstrate that incorporating the additional signal considerably reduces the bias in controlling the current, while the plasma shape errors are slightly higher but comparable. }
\label{fig:ssip}
\end{figure}

Continuous control trajectories typically include a \emph{transient} phase, where the agent state is rapidly changing towards achieving the target values, and a \emph{steady-state} phase where the agent is close enough to the desired target and control involves reacting to disturbances to remain close to the target. In traditional proportional-integral-derivative (PID) control, the policy includes linear feedback on the control error, its integral, and its derivative. The integral term in particular is designed to reduce/eliminate the bias in the steady-state error, while the derivative term helps to dampen the response to transient disturbances and reference changes. The feed-forward neural network policies used by \citet{degrave2022magnetic} do not use and cannot compute or construct an error integral, as the commanded actions are purely a function of the current inputs. An integral error approximation could be computed by a recurrent neural network, however, they have a greater risk of overfitting to the simulation dynamics. In this work, we implement a simpler solution: rather than having a policy learn the error integral, we manually compute it and append it to the set of observations seen by the feed-forward policy. We focus in particular on reducing the steady-state error in the plasma current ($I_p$), for which policies trained in \citet{degrave2022magnetic} exhibited significant bias and which can easily be computed. 
Diverging slightly from the traditional approach, we provide the network with the average plasma current error at time $t$, $e^{IF}_t$, defined as 
\begin{equation}
e^{IF}_t = \frac{1}{t} \sum_{i=1}^t e_i
\end{equation}
where $e_i$ is the difference between the plasma current measurement and reference values at time $i$. This choice of average keeps the numerical inputs better conditioned. Other choices are of course possible, for instance using the integrated error directly, or using an exponentially decaying average to put greater focus on the recent past.

We evaluate the benefits of incorporating the average error signal on the \shapetask task where the reference values for the plasma current and shape are constant, and the environment is initialized so that actual values are close to the references. Thus the main goal of the agent is to control the steady-state error.
Figure~\ref{fig:ssip} shows the simulated plasma current error trajectories for policies trained with and without the integrator feedback, with three random runs for each case. We note that the integrator feedback considerably reduces the plasma current bias, as expected.

\subsection{Episode Chunking}
\label{Episode Chunking}

\begin{figure}[ht!]
\centering
    \begin{subfigure}{0.68\linewidth}
        \centering
        \includegraphics[width=\textwidth]{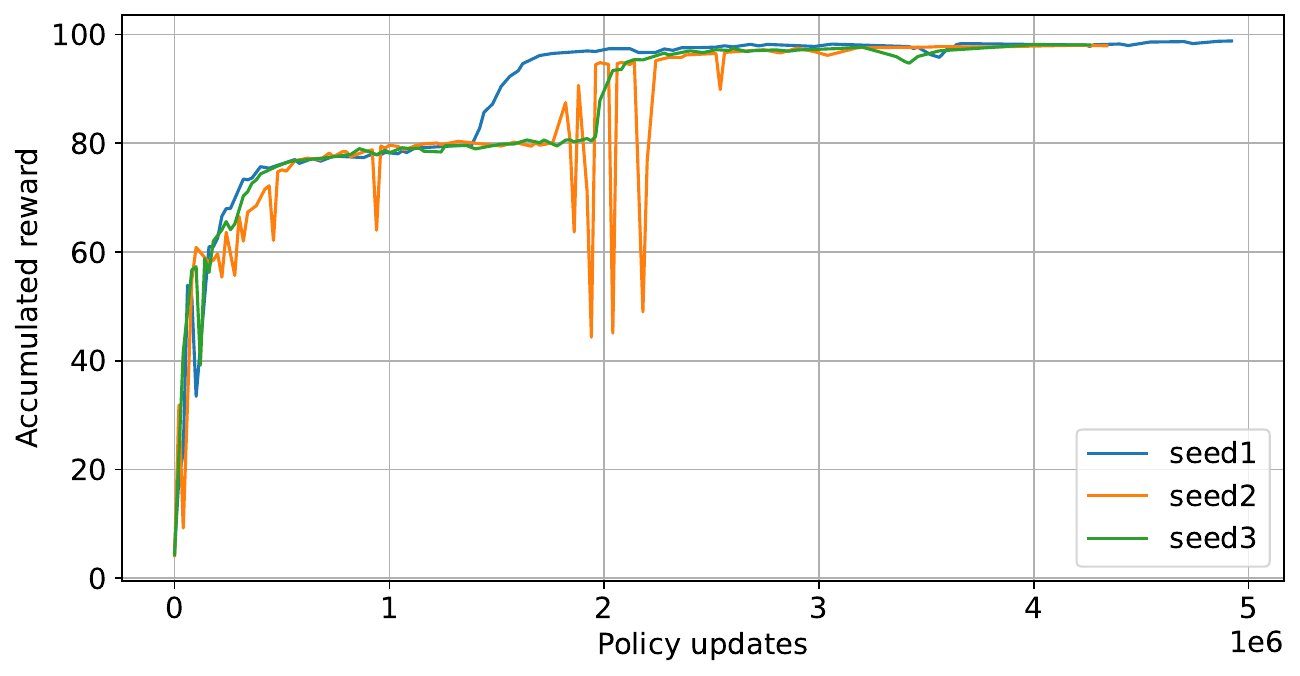}
        \caption{Learning curves for the "showcase\_xpoint" task for 3 different seeds.}
        \label{fig:showcase_xpoint_evaluator}
    \end{subfigure}
    % \vfill
    \begin{subfigure}{0.28\linewidth}
        \centering
        \includegraphics[height=0.25\textheight]{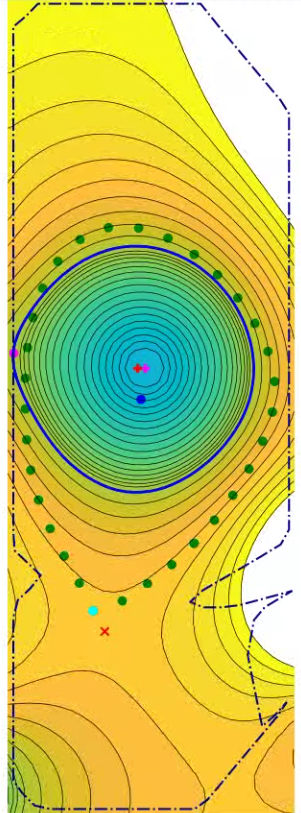}
        \caption{References and the actual plasma state during the first plateau phase.}
        \label{fig:showcase_xpoint_plateau_state}
    \end{subfigure}
\caption{RL agent learns the \showcasetask task in two distinct stages. (a) Smooth reward curve from 0 to 80 (out of 100), then a plateau and another smooth transition to the reward 100. (b) Before the transition LCFS (black solid line) is not being aligned with the references (green circles), which means that the agent knows how to hold the plasma but cannot generate the desired diverted shape.}
\label{fig:showcase_xpoint_learning_stages}
\end{figure}

The experiments on TCV last 1-2s, which corresponds to $10,000$ - $20,000$ time steps at the $10$kHz control rate. The FGE simulator \citep{carpanese2021development} (used to train the agents, as discussed above) takes around 2 seconds for a typical simulation step during training with stochastic actions on one core of AMD EPYC 7B12 CPU\footnote{The simulation step time varies from a fraction of a second to several seconds, depending on the plasma configuration. In contrast to deterministic policies, during training the RL agents explore the action space stochastically. This causes slower simulator convergence due to worse initial guesses for the iterative solver and due to encountering  plasma states on the limits of physical plausibility. As a result, the simulation in the RL setting is slower than in simple evaluation cases.}. Thus, FGE generates an episode with 10,000 steps in approximately 5 hours. This means that in the best case scenario, when the agent knows the best policy before the first trial, the training time would still be $\sim$5 hours (to observe the high-quality result).

In practice, RL agents need to explore the action space to find the best policy. Thus, depending on the task complexity, training times vary from days to weeks. Moreover, our tasks are structured such that the agent needs to learn somewhat independent ``skills'' sequentially. For example, in the \showcasetask task, the agent must first elongate the plasma, then shift its vertical position, then divert it and finally restore the original shape (see Fig. \ref{fig:showcase_xpoint_references}). 
We observe that learning for this task happens in two distinct stages (see Fig. \ref{fig:showcase_xpoint_evaluator}). First, the agent learns to manipulate limited plasmas, understanding how to elongate, move and hold the plasma, which corresponds to the smooth reward curve from 0 to around 80. At this stage, it is trying (but failing) to generate a diverted shape, instead it obtains the round LCFS with an inactive X-point, as demonstrated on Fig. \ref{fig:showcase_xpoint_plateau_state}. The reward plateaus at this level, until finally, the agent discovers how to successfully divert the plasma, where we see a transition in reward from 80 to near 100. In our standard setup, it takes several days of training for the agent to discover this transition.

\begin{figure}[ht!]
\centering
\includegraphics[width=0.15\linewidth]{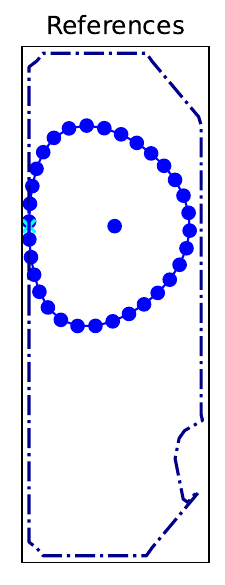}
\includegraphics[width=0.15\linewidth]{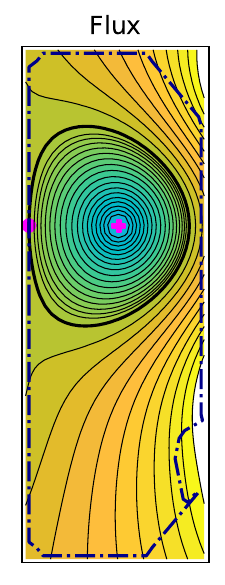}
\includegraphics[width=0.15\linewidth]{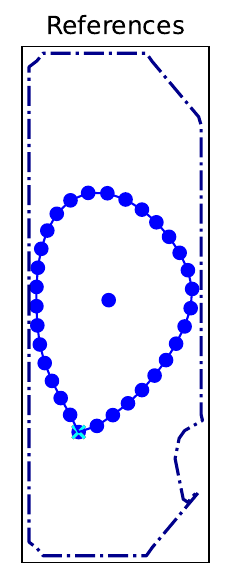}
\includegraphics[width=0.15\linewidth]{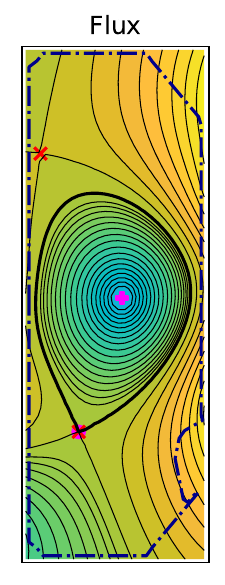}
\caption{FBT takes shape references and reconstructs the entire plasma state. Two different plasma shapes on the left are constructed via FBT and shown on the right. The initial state is visualized via the isolines of the magnetic flux.}
\label{fig:fbt-refs-plasma-state}
\end{figure}

This plateau lasts for an extended period because it is hard to learn to divert the plasma. This is compounded by the fact that the agent needs to divert at the specified time, so the right exploration needs to happen at the right time. On top of this, the long sequential episodes require the agent to go through the entire initial discharge phase before attempting to divert the plasma. On a slow simulator, these effects compound to create training times ranging from days for simple tasks to weeks for more complicated ones. Furthermore, \cite{degrave2022magnetic} shows that increasing the number of actors yields diminishing returns, so one cannot speed up training further by simply using more resources.

Our observations above (see Fig. \ref{fig:showcase_xpoint_learning_stages} and the related discussion) suggest that training can be accelerated by splitting the single long episode into a set of shorter episodes (``chunks'') and letting individual actors explore each chunk separately from the very beginning of training. This effectively parallelizes exploration. Fig. \ref{fig:showcase_xpoint_chunks} demonstrates different training setups with two and three chunks\footnote{Note that chunks in our setup can have different lengths.} for the \showcasetask task. To achieve this in practice we divide actors
%(typically the total of 1000)
into groups, where one group trains on the full-length discharge, and the other groups are each assigned a chunk. Each group starts the episode with the simulator in a state relevant to the selected chunk. The agent gets many more attempts at exploration from the shorter episodes. The specific division for any task is a hyperparameter; in general we have found success assigning 25-50\% of actors to the full episode and dividing the rest equally among the chunks. More complex schemes could also be employed, for instance adjusting the number of actors based on difficulty.
The simulator state at the beginning of each episode is specified by a given plasma shape (as on Fig. \ref{fig:showcase_xpoint_references}) and is computed using the FBT algorithm \cite{hofmann1988fbt}, which calculates the poloidal field coil currents needed to achieve a given plasma shape. This calculation neglects eddy current effects as well as time-varying Ohmic coil currents required to induce the plasma current, hence it is not guaranteed that this initial condition matches the end point of the previous chunk leading to chunk ``discontinuities''. Fig. \ref{fig:fbt-refs-plasma-state} demonstrates how two desired plasma shapes on the left are constructed with FBT and presented on the right. Plasma states are visually represented via the isolines of the magnetic flux (lines at which the magnetic flux is constant). We also display the last closed flux surface (solid black line), X-points (\textcolor{red}{red} crosses) and the limiter points (\textcolor{magenta}{magenta} circles).

In theory, misalignment between the tokamak state at the end of one chunk and the starting state at the next chunk (i.e. discontinuities) could create problems. This is particularly, but not exclusively, worrisome with regards to the control coil currents. For example, the agent may learn to aggressively ramp the coils, as coil currents ``reset'' between chunks and so risk of current saturation is reduced. Alternatively, the agent may learn to produce large swings in coil currents to ``smooth-out'' these discontinuities, which is undesirable and could lead to problems not necessarily captured in our model. In practice, we do not observe this to be a problem, as shown below. The agent learns to naturally resolve these discontinuities through training on the full episode, and so we do not see discontinuities in simulation in practice. This issue could also be resolved through more principled strategies, for example by widening the distribution of starting states for the chunks to provide greater overlap.

\begin{figure}[ht!]
\centering
\includegraphics[width=0.68\linewidth]{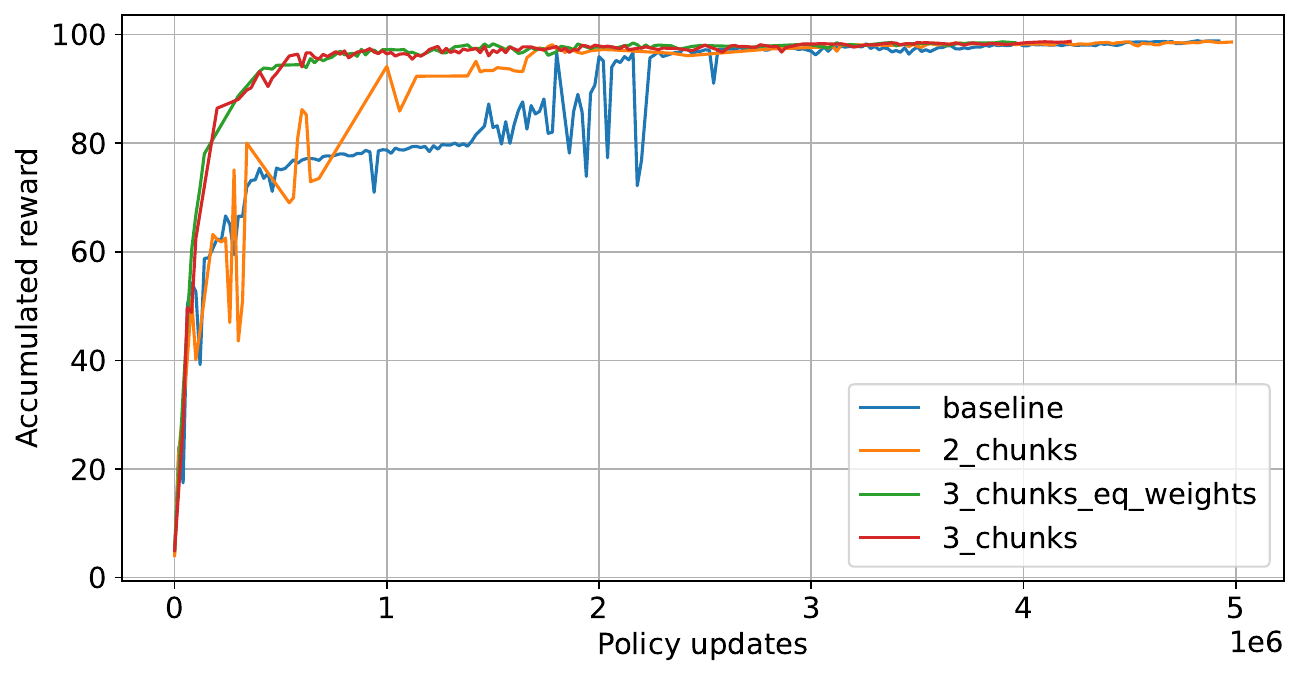}
\caption{Results for episode chunking applied to the \showcasetask task. Two and three-chunked setups are not only faster than the baseline (\textcolor{blue}{default}), but also have smoother learning curves. The two three-chunked setups considered are: equal actor distribution among all chunks (\textcolor{ForestGreen}{3\_chunks\_eq\_weights}) and more actors for the whole episode (\textcolor{red}{3\_chunks}). In all experiments the results are averaged over three random seeds.}
\label{fig:showcase_xpoint_chunks_results}
\end{figure}

Application of the chunking technique to the \showcasetask task with two/three chunks (depicted on Fig. \ref{fig:showcase_xpoint_chunks}) leads to significantly faster training times, as shown on Fig. \ref{fig:showcase_xpoint_chunks_results}. The two-chunked setup (\textcolor{orange}{orange} curve) is already faster than the baseline (\textcolor{blue}{blue}). The three-chunked setups (\textcolor{red}{3\_chunks} and \textcolor{ForestGreen}{3\_chunks\_eq\_weights}) not only provide further training speed up but also a much smoother learning curve. The agent achieves the reward of 96 (out of 100) in $\sim$ 10 hours compared to 40 hours for the baseline. Here we try two different three chunked setups: all actors are split in equally sized-groups (\textcolor{ForestGreen}{3\_chunks\_eq\_weights}); three times more actors are used for the whole episode compared to each other chunk. Both setups give similar results.

\subsection{Transfer learning}
\label{Transfer learning}
When attempting to reduce training time, a natural question is to ask if training from previous discharges can be re-used, that is, to what extent does the knowledge accumulated by the agent while solving an initial task transfer\footnote{See \citet{zhu2020transfer} for a survey of transfer learning in RL.} to a related target task.

Tokamak operators often experiment with different variations around a base task. Thus, we examine the transfer learning question when the target task is a variation on the initial task. Specifically, we examine performance when adjusting the reference plasma current and also shifting the location of the plasma.

We examine the performance of transfer learning in two forms
\begin{enumerate}
    \item Zero-shot: We run the policy learned for the initial task on the target task, without any additional data collection or policy parameter updates.
    \item Fine tuning: We initialize the policy and value function with the weights of the model learned on the original task, and then use those weights to train on the adjusted task through interacting with the environment with the new task as the reward. Note that this requires the same architecture to be used for both tasks (actor and critic network).
\end{enumerate}
In both cases, we use the parameters from an agent trained on the \showcasetask task as the initial parameters for the transfer.

\begin{figure}[h]
    \centering
    \begin{subfigure}{0.3\linewidth}
        \centering
        \includegraphics[width=\linewidth]{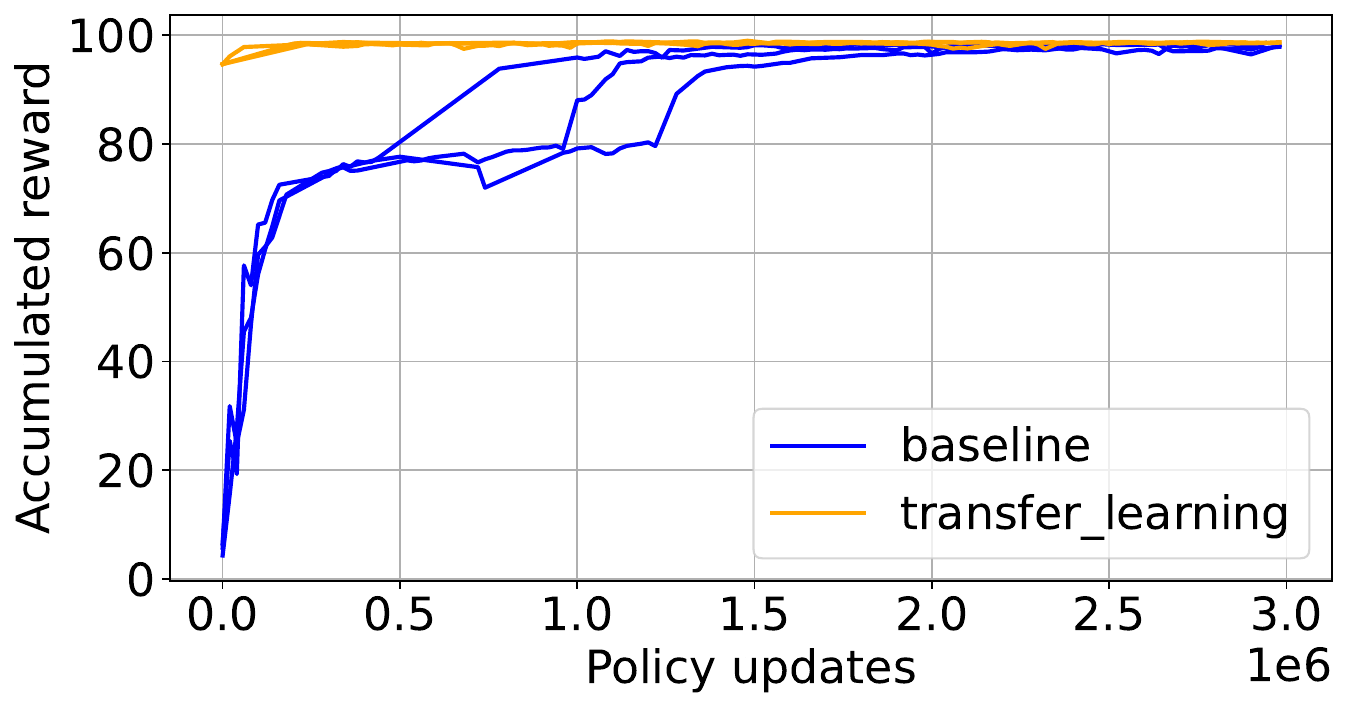}
        \caption{Increase target $I_p$ by 10kA.}
        \label{fig:showcase_xpoint_ip_shift_10ka}
    \end{subfigure}
    \begin{subfigure}{0.3\linewidth}
        \centering
        \includegraphics[width=\linewidth]{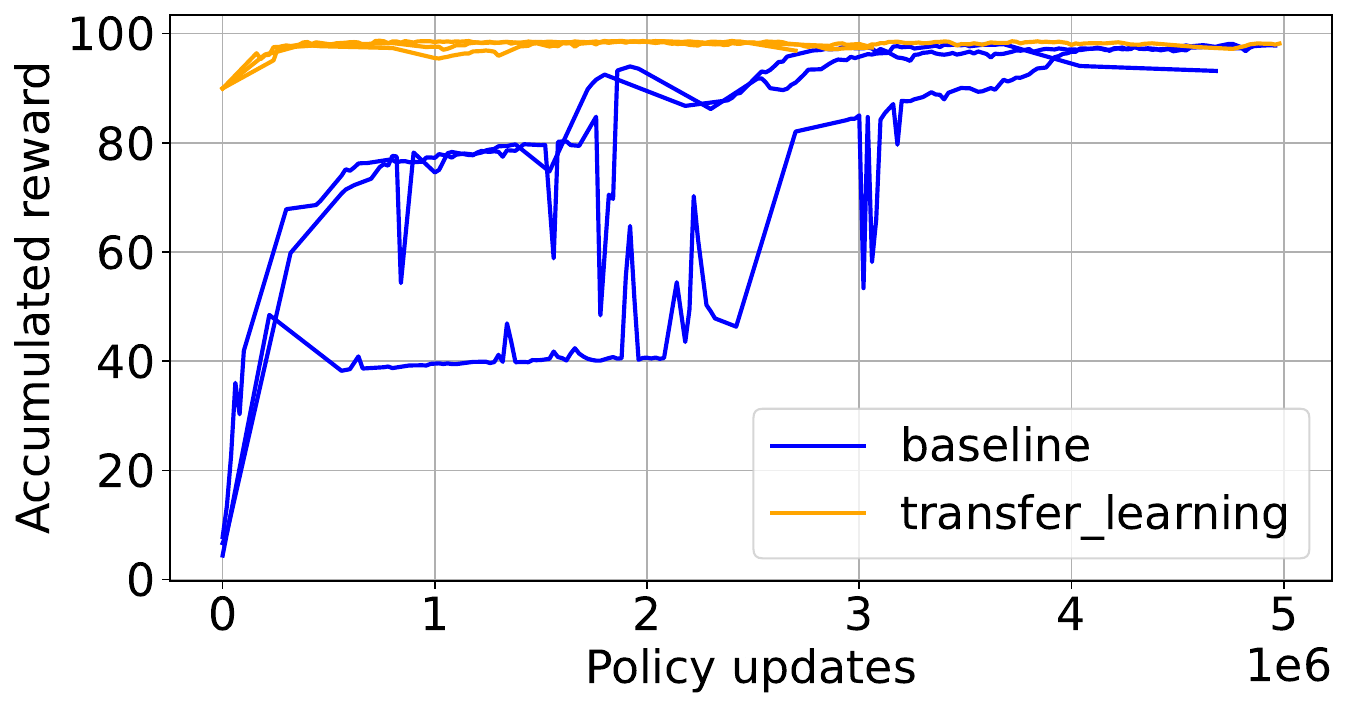}
        \caption{Increase target $I_p$ by 20kA.}
        \label{fig:showcase_xpoint_ip_shift_20ka}
    \end{subfigure}
     \begin{subfigure}{0.3\linewidth}
        \centering
        \includegraphics[width=\linewidth]{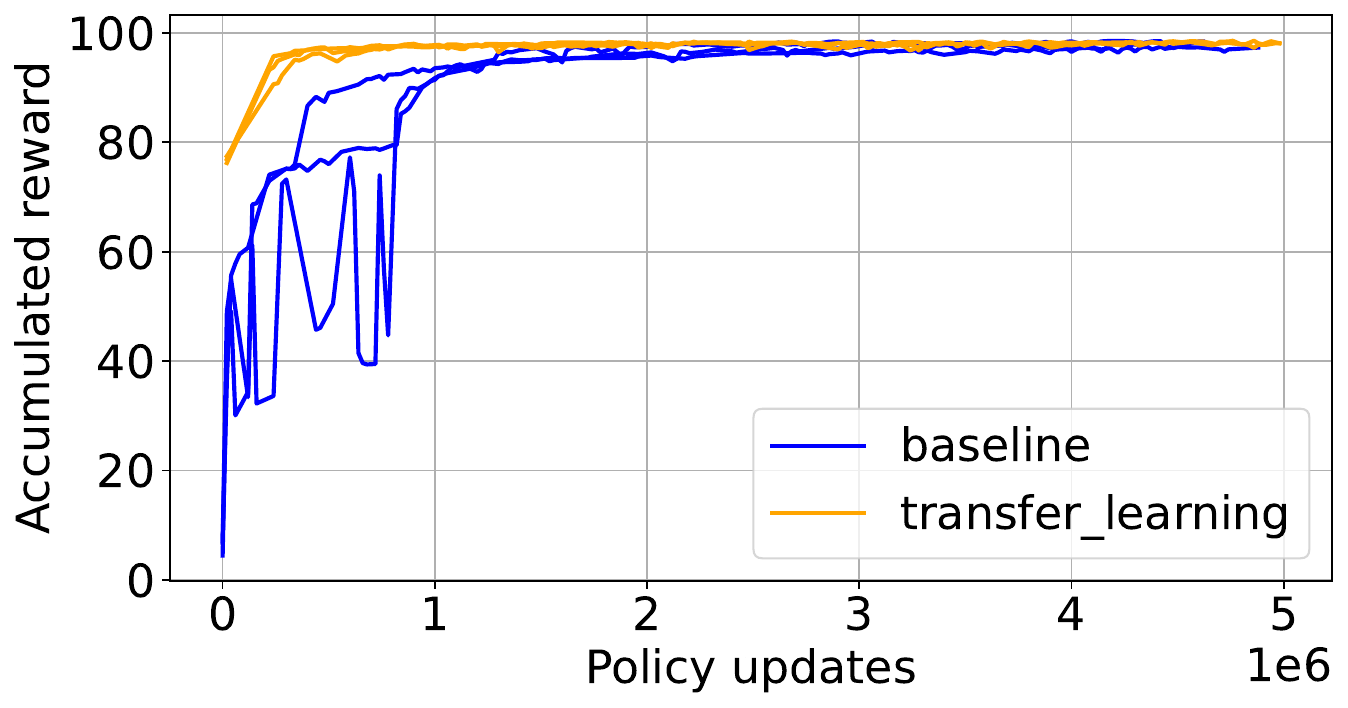}
        \caption{Decrease target $I_p$ by 50kA.}
        \label{fig:showcase_xpoint_ip_shift_50ka}
    \end{subfigure}
    \begin{subfigure}{0.3\linewidth}
        \centering
        \includegraphics[width=\linewidth]{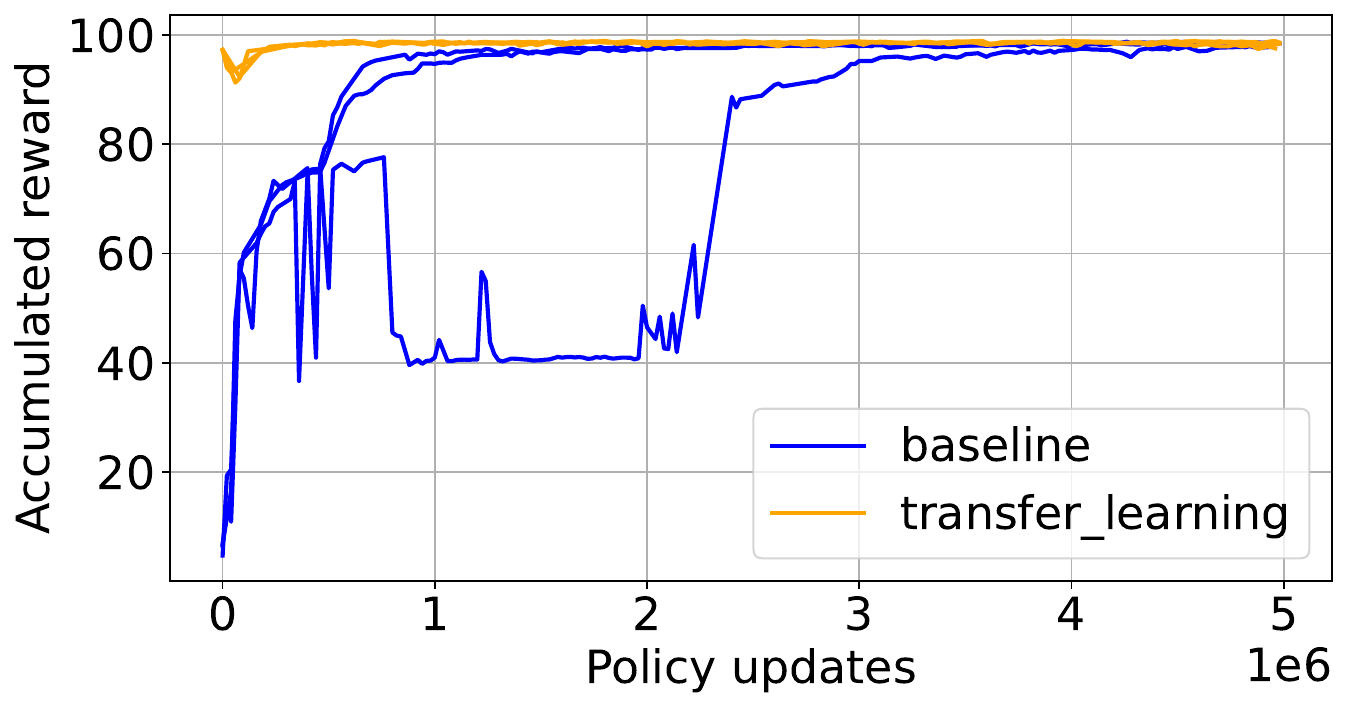}
        \caption{Shift target shape by 2cm.}
        \label{fig:showcase_xpoint_position_shift_2cm}
    \end{subfigure}
    \begin{subfigure}{0.3\linewidth}
        \centering
        \includegraphics[width=\linewidth]{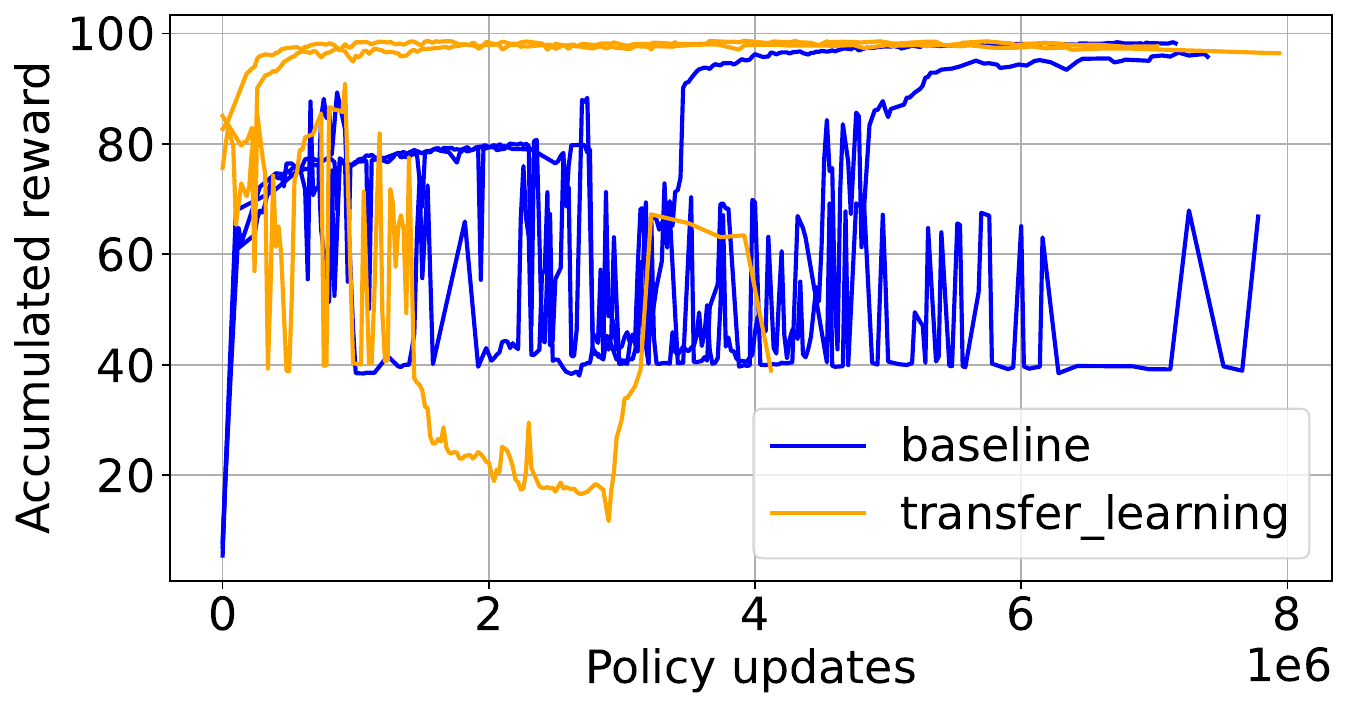}
        \caption{Shift target shape by 10cm.}
        \label{fig:showcase_xpoint_position_shift_10cm}
    \end{subfigure}
    \begin{subfigure}{0.3\textwidth}
        \centering
        \includegraphics[width=\textwidth]{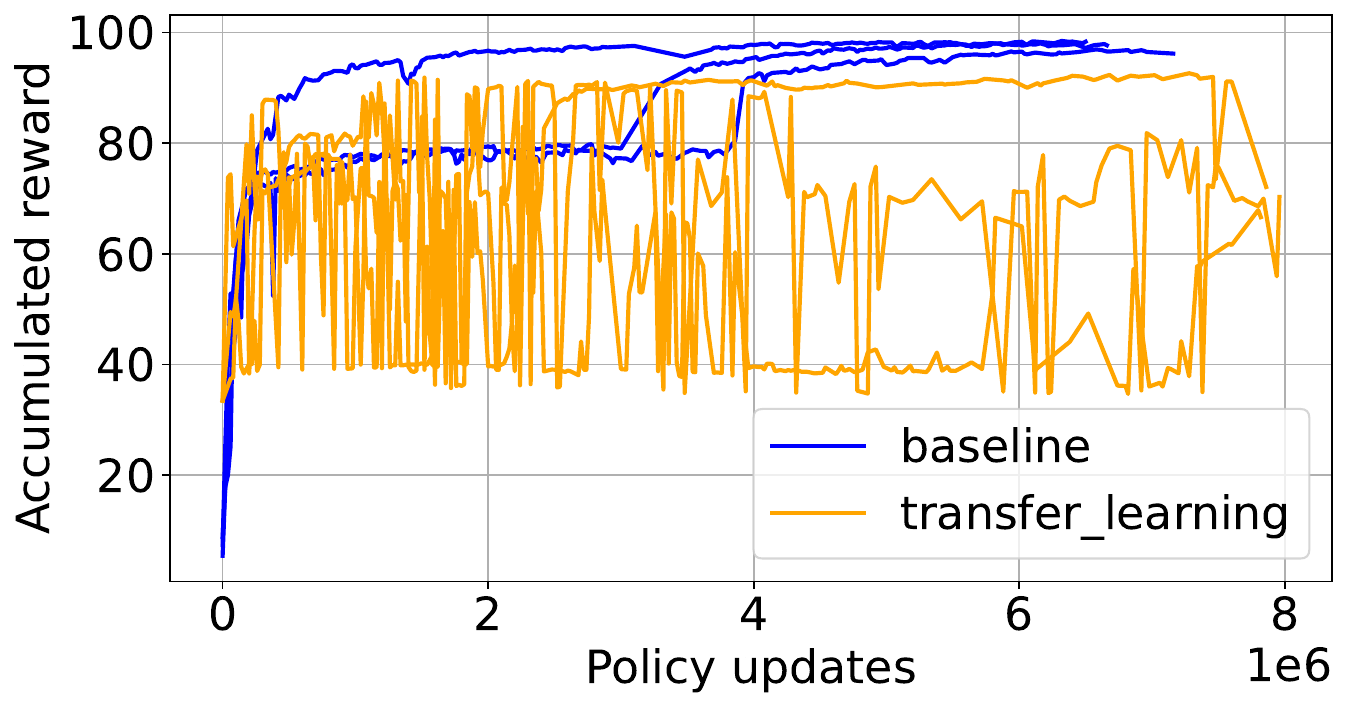}
        \caption{Shift target shape by 20cm.}
        \label{fig:showcase_xpoint_position_shift_20cm}
    \end{subfigure}
\caption{Fine tuning results from \showcasetask to two target tasks with different reference current (a, b, c) and vertical plasma position (d, e, f). The plots show total reward computed by the deterministic evaluator. The \textcolor{Blue}{blue} lines correspond to training on the target task \textcolor{Blue}{baseline}, while the \textcolor{Orange}{orange} lines correspond to \textcolor{Orange}{transfer\_learning} by initializing the agent with the model trained on \showcasetask. All experiments are run three times with different seeds.}
\label{fig:showcase_xpoint_transfer_learning}
\end{figure}

In the first experiment, we examine transfer when the reference  plasma current is adjusted to new reference levels. Concretely, we choose three variations where the target $I_p$ is adjusted from the baseline $-150$kA to first $-160$kA, then $-170$kA, and finally $-100$kA (specifically, we adjust the reference current in all time slices in Figure \ref{fig:showcase_xpoint_references} except for the initial handover level and the final rampdown level). We test the policy trained on \showcasetask, first without any additional training on the target task, and then allowing new training on the target task. The zero-shot results for reward and $I_p$ error are shown in \ref{table:zero_shot_ip}, where we see the agent does well for small changes to $I_p$, though struggles for larger shifts, in particular diverting the plasma for larger shifts.  The fine tuning results can be seen in Fig. \ref{fig:showcase_xpoint_ip_shift_10ka}, \ref{fig:showcase_xpoint_ip_shift_20ka}, \ref{fig:showcase_xpoint_ip_shift_50ka} and show that the fine tuning agents converge to a near-optimal policy faster than agents trained from scratch in all cases, although the difference is less pronounced for the largest $50kA$ shift.

\begin{table}[]
\centering
\begin{tabular}{|l|l|l|l|l|l|}
\hline
\textbf{$I_p$ target change}  & \textbf{Zero-shot reward} & \textbf{$I_p$ error (kA)}  \\ \hline
Baseline & 98.5 & 0.5 \\
\hline
$10$kA increase & 94.7 & 2.8 \\
\hline
$20$kA increase & 89.9 & 5.7 \\
\hline
$50$kA decrease & 76.7 & 6.3 \\
\hline
\end{tabular}
\caption{Reward for and $I_p$ error for zero-shot transfer with changes to the plasma current targets on the \showcasetask task. Values represent changes magnitude; a change from $-150$kA to $-160$kA is a $10$kA increase. The maximum reward for this task is $100$. The $I_p$ tracking error is computed during the phase of constant $I_p$ (0.25s to 0.85s). }
\label{table:zero_shot_ip}
\end{table}

We see that transfer learning is, in general, very effective when modifying the target plasma current. For small shifts, the unadjusted baseline agent performs almost as well as a specially trained agent.  As the adjustments to the current get larger, the zero-shot performance suffers, but we see that performance can be recovered with a small amount of fine-tuning. Assuming these results generalize, it implies a small range of plasma currents can be tested with the same base agent, and a larger range can be quickly trained from the base agent.

The second experiment examines variations in the plasma target location. Specifically, we adjust the target shape downward along the z axis, shifting by 2cm, 10cm and 20cm. For this experiment we observed the following:
\begin{enumerate}
    \item Zero-shot: Results are shown in Table \ref{table:zero_shot_shape}. We see that the zero-shot transfer works really well for the smallest shift (2cm), achieving over 97\% of best achievable performance of 100 for the task and a small shape error. For the larger 10cm shift, the performance is mediocre, seeing a reward of only 85, with a much larger error in the shape location. For the largest shift (20cm), the performance is poor, seeing a reward of only 35, due to a failure to successfully divert the plasma.
    \item Fine tuning: The fine tuning results can be seen in Fig. \ref{fig:showcase_xpoint_position_shift_2cm}, \ref{fig:showcase_xpoint_position_shift_10cm}, \ref{fig:showcase_xpoint_position_shift_20cm}, and show that transfer learning is clearly effective for the 2cm shift, and effective for two out of three seeds for the 10cm shift. For the larger 20cm shift, transfer learning seems detrimental rather than beneficial.
\end{enumerate}
Like with plasma current, we see that the baseline agent is capable of performing well for small adjustments to the shape. This similarly provides promise that an agent trained for a single condition can actually perform well for a variety of small adjustments without further training. However, the story is more mixed for fine-tuning, where using a baseline agent dramatically speeds up training for small adjustments, but is harmful for larger adjustments.

\begin{table}[]
\centering
\begin{tabular}{|l|l|l|l|l|l|}
\hline
\textbf{Shape target shift}  & \textbf{Zero-shot reward} & \textbf{Shape RMS error (cm)}  \\ \hline
Baseline & 98.5 & 0.67 \\
\hline
$2$cm & 97.3 & 0.78 \\
\hline
$10$cm & 84.7 & 3.4 \\
\hline
$20$cm & 33.5 & 4.9 \\
\hline
\end{tabular}
\caption{Reward and shape error for zero-shot transfer shifting the location of the plasma on the \showcasetask task. Values represent shifts downward in the domain. The maximum reward for this task is $100$. The shape RMS error is computed between 0.25s to 0.85s.}
\label{table:zero_shot_shape}
\end{table}

Overall, the results indicate that transfer learning can be useful, but also has limitations in its current form. As expected, the further the target task is from the initial task, the more the performance of transfer learning degrades, particularly for zero-shot learning. However, we note that it is relatively low cost (in CPU-hours) to run a simulated zero-shot evaluation to test performance before running a hardware experiment.  We also observed that some types of task changes allow for easier transfer than others - in our experiments, relatively large $I_p$ shifts seemed more suited to transfer learning than large position shifts, which is understandable given the relative complexity of the tasks. Further study is needed to understand which tasks are amenable to transfer learning, and how to expand the regions of effective transfer, both zero-shot and fine tuning.

\section{Tokamak Discharge Experiments on TCV}

The previous sections have focused solely on simulation, training and assessing control policies with the FGE simulator. Given the complexity and challenges of Tokamak modeling, it is important to not blindly accept performance improvements in simulation as identical to performance improvements in physical discharges. While better simulation results might be \emph{necessary} for improved results on the actual Tokamak, they are not always \emph{sufficient} and model mismatch errors may begin to dominate without additional explicit work to reduce sim-to-real gaps. 
This is especially the case for policies obtained using RL which are known to overfit to imperfect simulators \citep{zhang2018study}. We thus tested several of the aforementioned simulation enhancements on dedicated discharges on the TCV tokamak. This way we can assess the strengths and limitations of our current work, and provide direction for the next set of improvements. 

\subsubsection{Reward Shaping for Plasma Shape Accuracy}
We examine the accuracy improvements seen from reward shaping for two different configurations and objectives: reducing the LCFS error in a shape stabilization task and improving X-point accuracy for the \snowflaketask task configuration. We compare the results seen in simulation with those on TCV, and to comparable TCV experiments from \cite{degrave2022magnetic}. Like \cite{degrave2022magnetic}, we deploy the control policy by creating a shared-library object out of the actor network (defined by a JAX graph), where the commanded action is taken as the mean of the output Gaussian distribution.

We first test a control policy trained to reduce the LCFS error in the \textit{shape\_70166} stabilization task using the reward shaping approach discussed above in the \nameref{Reward Shaping} section.

For this stabilization task, we use TCV's standard breakdown procedure and initial plasma controller. At 0.45s, control is handed over to the learned control policy, which then tries to maintain a fixed plasma current and shape for a duration of $1$s. After the discharge, we compute the reconstructed equilibria using the LIUQE code  \citep{moret2015tokamak}. At each $0.1$ms slice during the $1$s discharge, we compute the error in the plasma shape.
We compare the accuracy from three experiments, measuring the shape error from a simulated discharge and a TCV discharge:
\begin{enumerate}[label=(\alph*)]
\item a baseline RL controller that pre-dates this work (``Previous''), 
\item an updated baseline agent using the updated training infrastructure used in this work (``Updated''),
\item a an agent trained using reward shaping, as in Fixed Reward describe in the \nameref{Reward Shaping} section.
\end{enumerate}
The results of these runs are reported in Table \ref{table:craquelin_experiment}.

Both recent policies, the updated baseline and reward-shaped policy, significantly outperform the pre-existing baseline for the goal of reducing LCFS error as shown in Table \ref{table:craquelin_experiment}. This reduction is due to improvements in the training infrastructure. These improvements are also seen on the TCV experiments --- both the updated baseline and the reward-shaped agent outperform the previous baseline. However, if we compare the two contemporary experiments, entitled Updated and Reward-Shaped respectively, we see that the shaped policy performs worse on TCV relative to the updated baseline despite achieving better results in simulation. One hypothesis for this difference is that the plasma resistivity during the TCV discharge was near the edge of the range of variation used during training. It is possible that the shaped controller was less robust to these variations. Over the course of the discharge, the difference in Tokamak state caused by the higher plasma resistivity (coil currents, etc.) compounds. This could explain the fact that the error is low during the initial stage of the discharge, but then grows in time as the discharge proceeds. Overall, we see that improvements to performance in simulation are beneficial, the accuracy of the updated infrastructure is higher that the previous baseline. However, there is a limit to optimizing simulation performance. Indeed, it seems that for this case, there is little to be gained by further reducing the simulation RMS error, and instead we should now focus on addressing the sim-to-real gap.

\begin{figure}
\centering
\includegraphics[width=100mm]{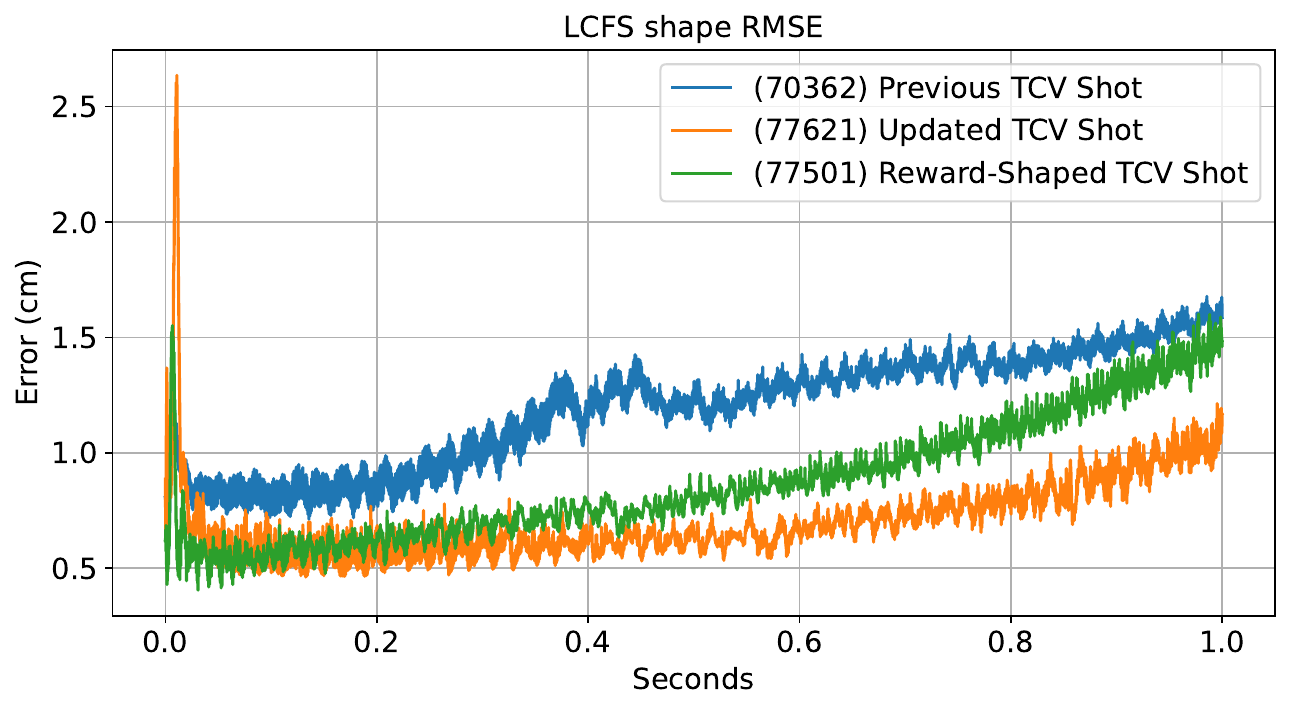}
\caption{LCFS error on \shapetask stabilization task.}
\label{fig:craquelin}
\end{figure}

\begin{table}[]
\centering
\begin{tabular}{|l|l|l|l|l|l|}
\hline
\textbf{Experiment}  & \textbf{LCFS Mean RMSE (cm)}  \\ \hline
Previous Simulation & 0.688 \\
\hline
Updated Simulation & 0.350 \\
\hline
Reward-Shaped Simulation & \textbf{0.178} \\
\hline
\hline
(70362) Previous TCV Shot &  1.209 \\
\hline
(77621) Updated TCV Shot & \textbf{0.698} \\
\hline
(77501) Reward-Shaped TCV Shot & 0.863 \\
\hline
\end{tabular}
\caption{Comparison of policies on LCFS shape error for the \shapetask stabilization task}
\label{table:craquelin_experiment}
\end{table}

\subsubsection{Reward Shaping for X-Point Location Accuracy}
\begin{figure}[h]
\centering
\includegraphics[width=\textwidth]{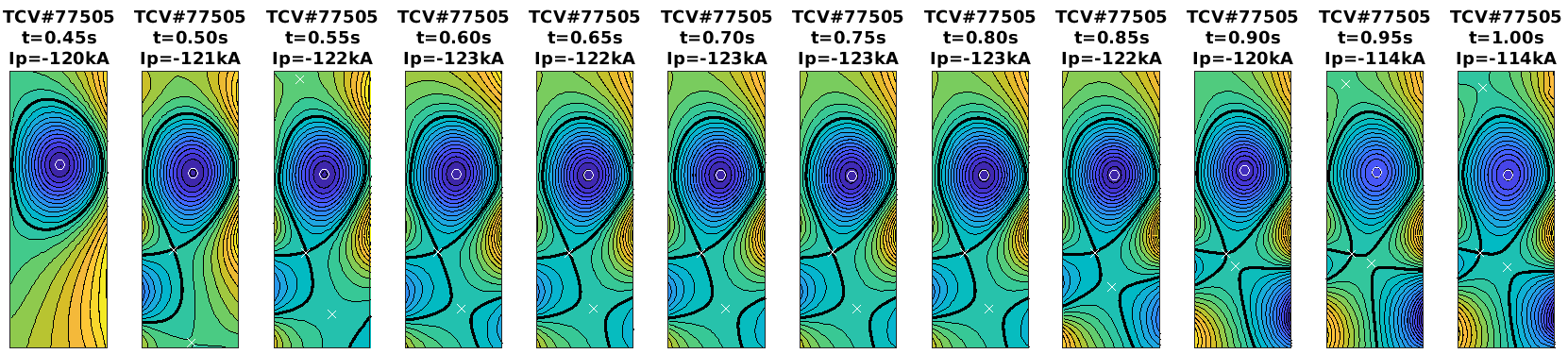}
\caption{Evolution of TCV equilibrium reconstructed post-shot using LIUQE based on magnetic measurements, for Snowflake TCV shot (77505).}
\label{fig:moondance_steps}
\end{figure}

We next compare the effects of reward shaping on the more complicated ``snowflake'' configuration, as shown in Figure \ref{fig:moondance_steps}. The training reward for this policy was shaped to increase the accuracy of X-point control. As in the stabilization experiment, the plasma is created and initially controlled by the standard TCV procedures, handing over to the RL controller at 0.45s. In this experiment, the RL-trained policy successfully established a snowflake with two X-points at a distance of $34$cm. The policy then successfully brought the two X-points to a targeted distance of $6.7$cm, coming close to establishing a so-called ``perfect snowflake''. However, at 1.0278s (0.5778s after handover), the plasma disrupted due to vertical instability. Upon inspection, it seems that the controller struggled to keep a consistent shape, where the vertical oscillations increased, and the active X-point switched between the two X-points, leading to a loss of control. 
Table \ref{table:moondance_experiment} shows the accuracy of X-point tracking during the window where the plasma was successfully controlled. The performance during this experiment is compared with the equivalent snowflake experiment reported in \cite{degrave2022magnetic}. Similar to above, we compute the error from the plasma states re-constructed by LIUQE. We see that the substantial improvements to X-point accuracy seen in simulation do indeed lead to significant improvements in X-point accuracy seen on hardware. 
The improvements from reward-shaping result in a 59.7\% reduction in RMSE tracking distance over the control window compared to a previous TCV experiment. Other metrics, such as the LCFS, report a minimal decrease in accuracy, which is expected, as described in \nameref{Reward Shaping}. Here, we do indeed see notable benefits from reward shaping, though work remains on bridging the sim-to-real gap for maintaining highly-accurate perfect snowflakes.

\begin{table}[]
\centering
\begin{tabular}{|l|l|l|}
\hline
\textbf{Experiment} & \textbf{X-point Mean RMSE}  & \textbf{LCFS Mean RMSE}  \\ \hline
Previous Simulation & 3.46cm & 0.078 \\ \hline
Reward-Shaped Simulation & \textbf{0.27cm} & \textbf{0.0069} \\ \hline
\hline
(70755) Previous TCV Shot & 3.66cm & \textbf{0.0110}  \\ \hline
(77505) Reward-Shaped TCV Shot & \textbf{1.74cm} & 0.0119  \\ 
\hline
\end{tabular}
\caption{Comparison of policies for X-point tracking for the Snowflake configuration}
\label{table:moondance_experiment}
\end{table}

\begin{figure}
\centering
\includegraphics[width=100mm]{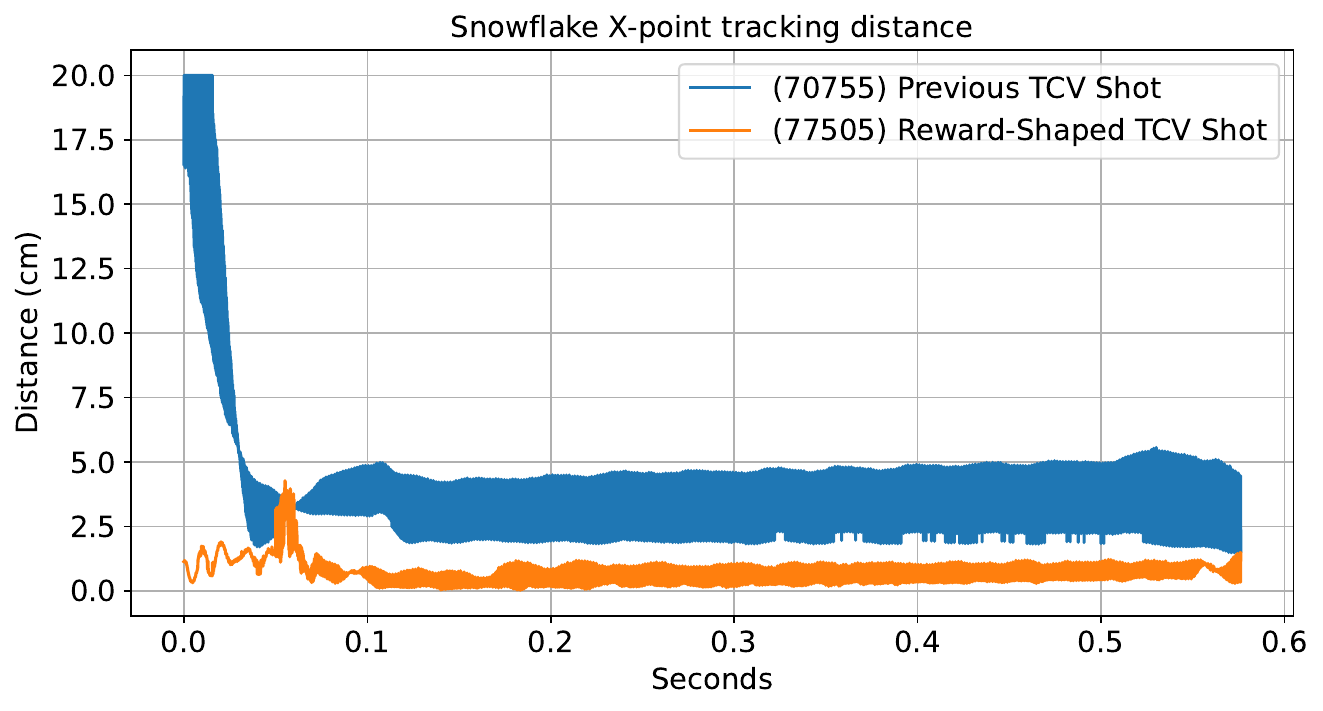}
\caption{X-point tracking distance RMSE comparison against a previous comparison TCV discharge.}
\label{fig:moondance_xpoint}
\end{figure}

\subsubsection{Validation of Accelerated Training via Episode Chunking}
\begin{figure}[h]
\centering
\includegraphics[width=\textwidth]{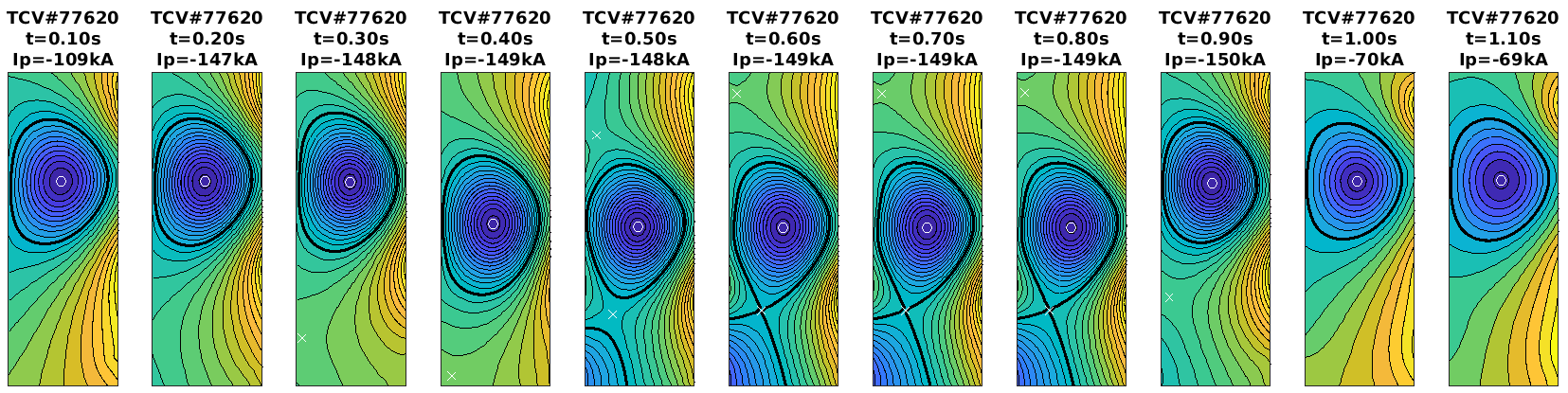}
\caption{Evolution of TCV equilibrium reconstructed post-shot using LIUQE based on magnetic measurements,  for showcase TCV shot (77620).}
\label{fig:metamorph_steps}
\end{figure}

Finally, we validate the use of \nameref{Episode Chunking} to reduce training time, especially to verify that the possible  ``discontinuities''  from episode chunking do not show up in TCV discharges. We ran an experiment for the showcase configuration trained using 3 chunks. The time-trace of reconstructed equilibria for this experiment can be seen in Figure \ref{fig:metamorph_steps} . We find that the experiment went as expected, with no noticeable artifacts due to the episode chunking. This demonstrates that there is no loss of quality from this training acceleration approach.

\section{Conclusions and Future Work}

There is excitement around the potential impact of reinforcement learning for magnetic control of Tokamaks, but drawbacks limit their uptake within the community. In this paper, we significantly improved upon several key limitations, with a focus on policy accuracy and overall training speed.

We first addressed the issue of controller accuracy. In the \nameref{Reward Shaping} section, we showed that reward shaping and tuning can significantly improve the controller accuracy, reaching a 65\% reduction in LCFS error in simulation. We further showed that providing an integral observation to the agent significantly reduces the long-term bias of the agent. Combined, these provide promise in the ability for RL to generate highly accurate controllers.

We then demonstrated the effectiveness of \nameref{Episode Chunking} to alleviate exploration challenges and help the agent discover control policies for complex configurations. Dividing training episodes in this way significantly reduced the training time for an example diverted plasma. We additionally showed that \nameref{Transfer learning} by warm-starting training can allow for the rapid generation of policies when minor adjustments are made to the training task. These are two powerful tools that provide significant reductions in the amount of time needed to train new policies.

While these results significantly reduce the limitations on reinforcement learning controllers, there is still a lot of room for improvement. Going forward, there will need to be increased focus not only on improving performance in simulation, but on matching that level of performance during actual plasma discharges on hardware. In particular, the experiments in Table \ref{table:moondance_experiment} show that the gap in accuracy between simulation and hardware is now close to dominating any remaining improvements in simulation.

There are a number of promising directions for improving hardware transfer, for example by improving the modeling of plasma parameter variation (to expose the agent to more realistic scenarios), and by improving the real-time knowledge of the agent, for example by integrating real-time plasma observers directly into the observations of the agent. More ambitiously, substantial improvement could be gained by using fine-tuning to update policies in response to experimental data. Such data could be used to directly fine-tune the weights of a policy for a specific experiment, or alternatively be used to improve simulation capability, thus indirectly improving agent quality. In either case,  this will be challenging given the paucity of data.

Similarly, there are many opportunities for continued reduction of the training time requirements. Our results on episode chunking suggest that exploration is a significant bottleneck to training time. Explicit exploration techniques, for example \cite{taiga2019benchmarking}, could overcome the bottlenecks in taking the right action at critical moments, and thus significantly reduce training time. Relatedly, using training data from previous experiments combined with offline RL approaches \citep{levine2020offline} could provide key demonstrations, avoiding the need for the agent to `unlock' each difficult moment anew during training. One could also use pre-computed feed-forward coil current trajectories from existing optimizers as a starting point for policy creation. 

Alternative model architectures to the existing MLPs and LSTMs could provide significant benefits. State-space models \citep{gu2021efficiently} are one promising approach to modeling long range dependencies without sacrificing inference speed. Another promising direction is looking into foundation models \citep{bommasani2021opportunities}, which have shown impressive generalization and fine-tuning capabilities \citep{brohan2022rt, team2023human}. Potentially, a single large-scale model could learn to control many plasma discharges, and adapt to specific scenarios after a few trials. One significant challenge for this direction, however, is generating a policy that can execute at the high frequencies required for plasma control.

Overall, reinforcement learning remains an attractive alternative for plasma control. This work has begun to alleviate some of the remaining blockers to adoption for the application of magnetic control, and there are many promising directions for continued enhancement.

\section{Acknowledgements}

We thank Anton Zhernov for help during the development of this project. We thank Kieran Milan and Jerry Luo for strategic help during the project.
This work was supported in part by the Swiss National Science Foundation.

This work has been carried out within the framework of the EUROfusion Consortium, via the Euratom Research and Training Programme (Grant Agreement No 101052200 — EUROfusion) and funded by the Swiss State Secretariat for Education, Research and Innovation (SERI). Views and opinions expressed are however those of the author(s) only and do not necessarily reflect those of the European Union, the European Commission, or SERI. Neither the European Union nor the European Commission nor SERI can be held responsible for them.

\bibliography{main}

\clearpage
\appendix
\section{Appendix}

\subsection{References for tasks and episode chunking}

\begin{figure}[ht!]
    \centering
    \begin{subfigure}{\linewidth}
        \centering
        \subcaption{The \showcasetask task.}
        \includegraphics[width=\linewidth]{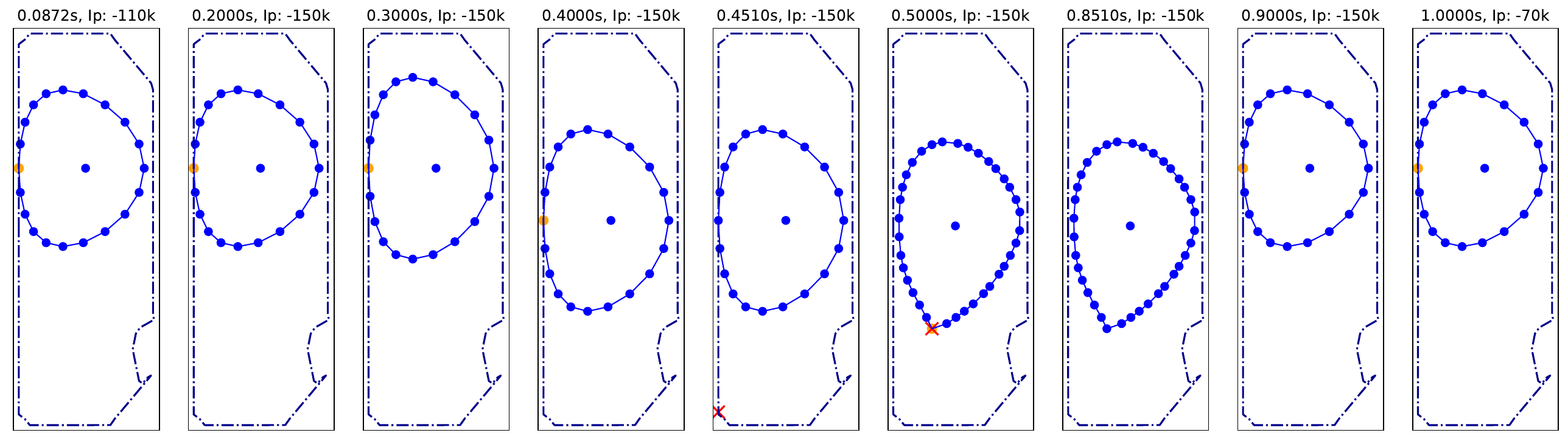}
        \label{fig:showcase_xpoint_references_all}
    \end{subfigure}
    \vfill
    \begin{subfigure}{\linewidth}
        \centering
        \subcaption{The \showcasetask task - two chunks.}
        \includegraphics[width=\linewidth]{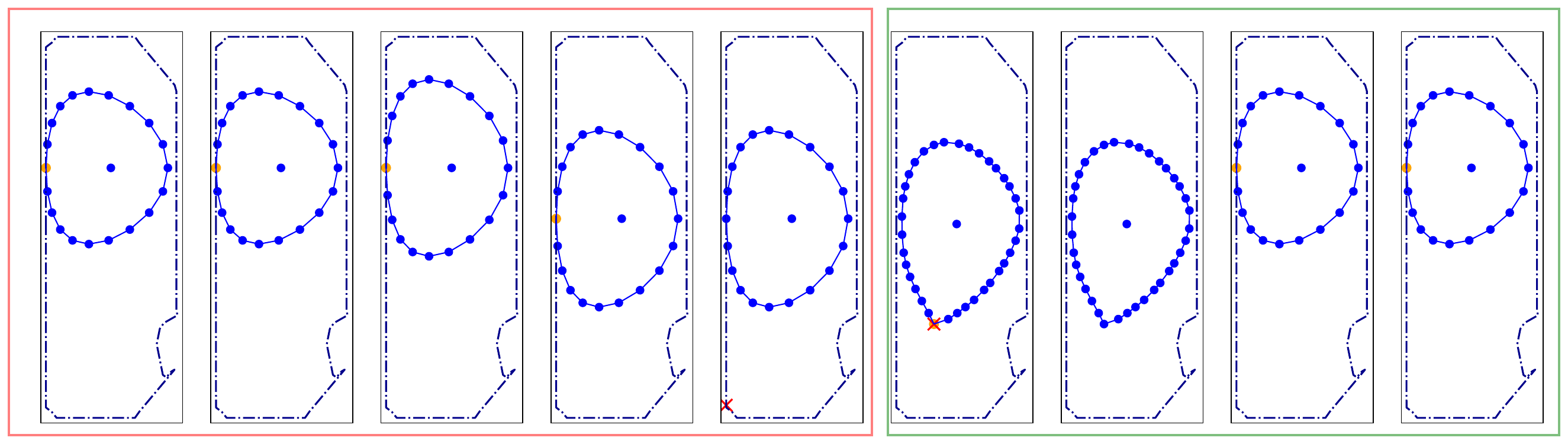}
        \label{fig:showcase_xpoint_references_2_chunks}
    \end{subfigure}
    \vfill
    \begin{subfigure}{\linewidth}
        \centering
        \subcaption{The \showcasetask task three chunks.}
        \includegraphics[width=\linewidth]{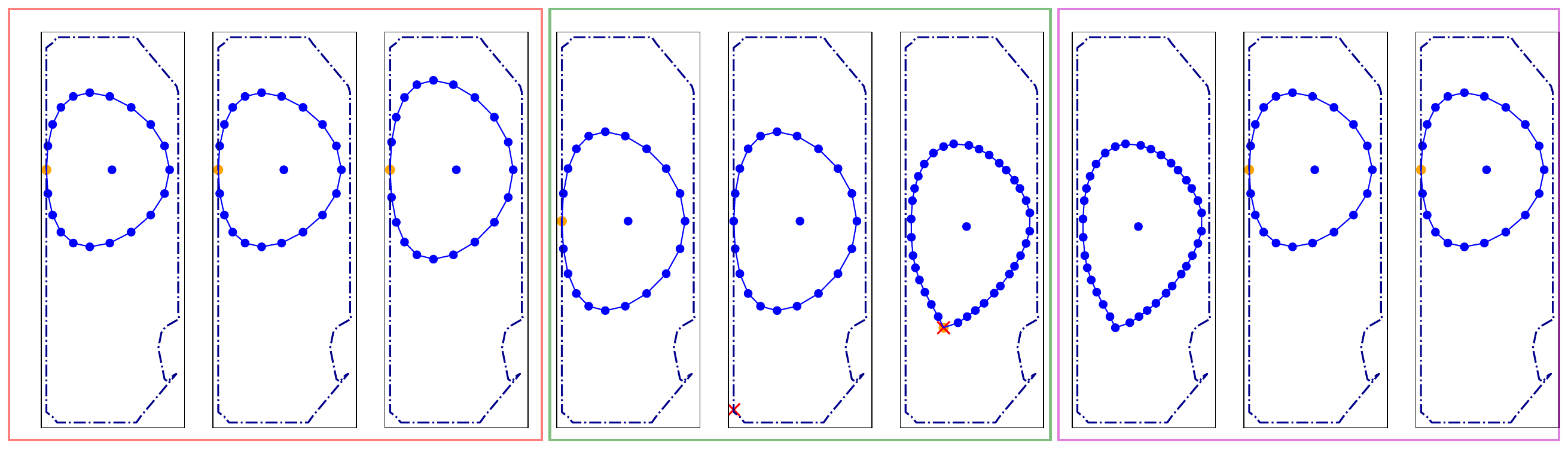}
        \label{fig:showcase_xpoint_references_3_chunks}
    \end{subfigure}
\caption{Example of different training setups for the \showcasetask task used in \nameref{Episode Chunking}. In the original setup (a) all actors start from the first shape on the left. In case of two chunks (b) some actors start from the most left shape and finish with the forth shape (\textcolor{red}{chunk 1}) after which the episode ends, and others start from the fifth shape (\textcolor{green}{chunk 2}). Analogously one can create three (c) and more chunks.}
\label{fig:showcase_xpoint_chunks}
\end{figure}

\begin{figure}[ht!]
\centering
\includegraphics[width=0.68\linewidth]{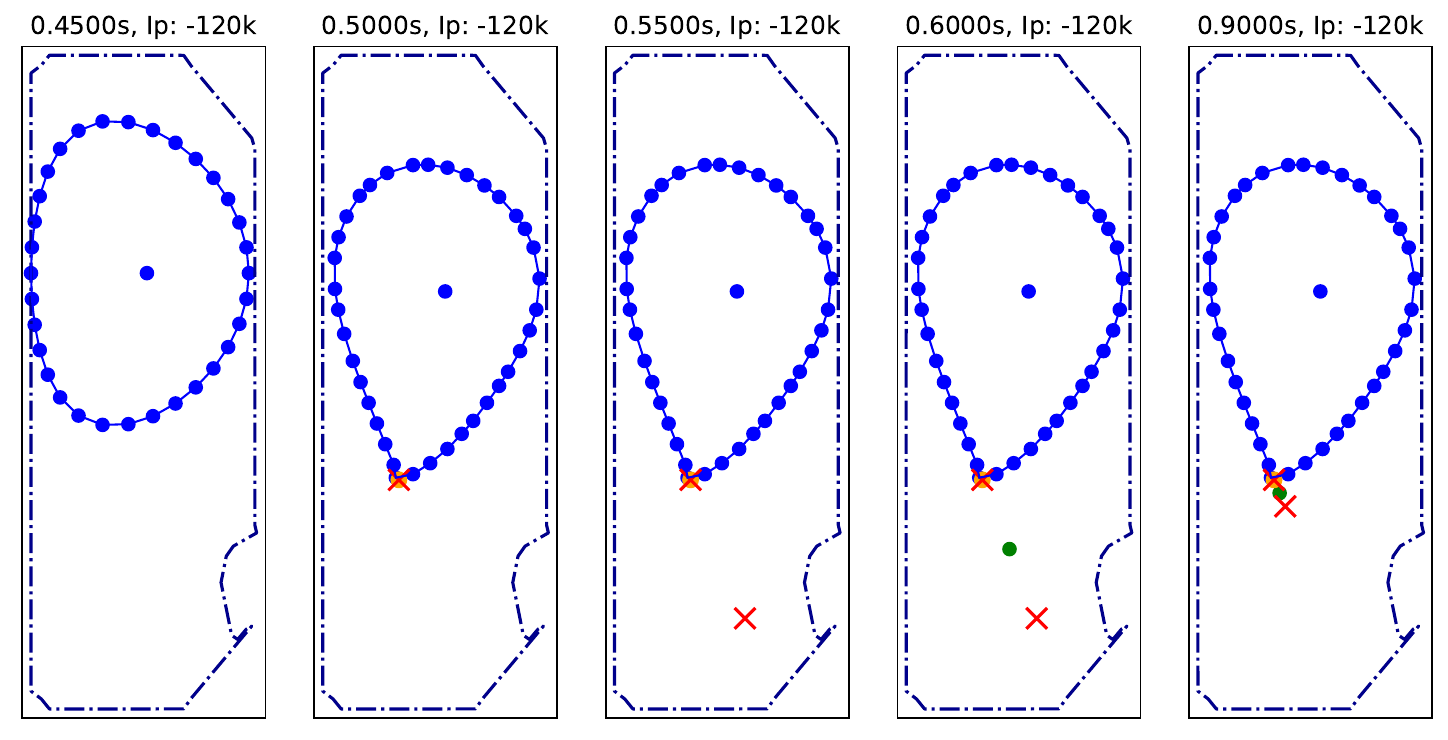}
\caption{References for the \snowflaketask task.}
\label{fig:snowflake_to_perfect_references}
\end{figure}

\end{document}